\documentclass[12pt]{article}
\usepackage{amsmath,amssymb,amsthm}
\usepackage{graphicx,subfigure}
\usepackage{changepage}
\usepackage{hyperref,cite,url,breakurl}
\usepackage{mathtools,xparse,MnSymbol}
\usepackage{multirow}
\usepackage{stackrel}
\begin{document}
\baselineskip 0.6cm

\def\bra#1{\langle #1 |}
\def\ket#1{| #1 \rangle}
\def\inner#1#2{\langle #1 | #2 \rangle}
\def\brac#1{\llangle #1 \|}
\def\ketc#1{\| #1 \rrangle}
\def\innerc#1#2{\llangle #1 \| #2 \rrangle}
\def\app#1#2{%
  \mathrel{%
    \setbox0=\hbox{$#1\sim$}%
    \setbox2=\hbox{%
      \rlap{\hbox{$#1\propto$}}%
      \lower1.1\ht0\box0%
    }%
    \raise0.25\ht2\box2%
  }%
}
\def\approxprop{\mathpalette\app\relax}
\DeclarePairedDelimiter{\norm}{\lVert}{\rVert}

\begin{titlepage}

\begin{flushright}
\end{flushright}

\vskip 0.6cm

\begin{center}
{\Large \bf Ensemble from Coarse Graining:\\ Reconstructing the Interior of an Evaporating Black Hole}

\vskip 0.7cm

{\large Kevin Langhoff$^{a,b}$ and Yasunori Nomura$^{a,b,c}$}

\vskip 0.5cm

$^a$ {\it Berkeley Center for Theoretical Physics, Department of Physics,\\
  University of California, Berkeley, CA 94720, USA}

\vskip 0.2cm

$^b$ {\it Theoretical Physics Group, Lawrence Berkeley National Laboratory, 
 Berkeley, CA 94720, USA}

\vskip 0.2cm

$^c$ {\it Kavli Institute for the Physics and Mathematics of the Universe (WPI),\\
 UTIAS, The University of Tokyo, Kashiwa, Chiba 277-8583, Japan}

\vskip 0.8cm

\abstract{In understanding the quantum physics of a black hole, nonperturbative aspects of gravity play important roles. In particular, huge gauge redundancies of a gravitational theory at the nonperturbative level, which are much larger than the standard diffeomorphism and relate even spaces with different topologies, allow us to take different descriptions of a system. While the physical conclusions are the same in any description, the same physics may manifest itself in vastly different forms in descriptions based on different gauge choices.

In this paper, we explore the relation between two such descriptions, which we refer to as the global gauge and unitary gauge constructions. The former is based on the global spacetime of general relativity, in which understanding unitarity requires the inclusion of subtle nonperturbative effects of gravity. The latter is based on a distant view of the black hole, in which unitarity is manifest but the existence of interior spacetime is obscured. These two descriptions are complementary. In this paper, we initiate the study of learning aspects of one construction through the analysis of the other.

We find that the existence of near empty interior spacetime manifest in the global gauge construction is related to the maximally chaotic, fast scrambling, and universal dynamics of the horizon in the unitary gauge construction. We use the complementarity of the gauge choices to understand the ensemble nature of the gravitational path integral in global spacetime in terms of coarse graining and thermality in a single unitary theory that does not involve any ensemble nature at the fundamental level. We also discuss how the interior degrees of freedom are related with those in the exterior in the two constructions. This relation emerges most naturally as entanglement wedge reconstruction and the effective theory of the interior in the respective constructions.}

\end{center}
\end{titlepage}

\section{Introduction and Grand Picture}
\label{sec:intro}

A black hole is a special object.
When described in the global spacetime of general relativity, it has a spacetime region from which nothing can escape to spatial infinity without encountering a singularity.
When viewed from a distance, the large redshift caused by a strong gravitational field makes the density of states exponentially large even at the quantum level~\cite{Bekenstein:1973ur,Hawking:1974sw}, while at the same time a classically diverging redshift factor makes the intrinsic scale of dynamics reach the string scale near the horizon, causing the breakdown of the classical spacetime description~\cite{Susskind:1993if}.
This string scale dynamics seems to exhibit a variety of universal behaviors as explored in Refs.~\cite{Maldacena:2015waa,Hayden:2007cs,Sekino:2008he,Banks:2010zn,Harlow:2018jwu}.

These two seemingly different pictures, however, can be different manifestations of the same physics~\cite{Nomura:2018kia,Nomura:2019qps,Nomura:2019dlz}.
The basic idea is that the two pictures give equivalent physical conclusions after taking into account huge gauge redundancies of gravitational theory at the nonperturbative level~\cite{Marolf:2020xie,McNamara:2020uza}, which are much larger than the standard diffeomorphism and relate even spaces with different topologies~\cite{Marolf:2020xie,Jafferis:2017tiu}.
The special features of the ultraviolet (UV) string dynamics manifest in one way of fixing the redundancies are related with the existence of large interior spacetime in the infrared (IR) manifest in another fixing~\cite{Nomura:2018kia,Nomura:2019qps}.
The two constructions are complementary.
In the construction based on global spacetime, one can use semiclassical techniques to understand certain important aspects of the black hole physics~\cite{Penington:2019npb,Almheiri:2019psf,Almheiri:2019hni,Penington:2019kki,Almheiri:2019qdq}.
In the other construction based on a distant view, one may find a restriction on the applicability of semiclassical theory based on the (non-)existence of approximate linear operators representing the excitations~\cite{Nomura:2019dlz}.

The gist of this paper is to explore the relation between the two constructions described above and initiate the study of learning aspects of one construction through the analysis of the other.
From the quantum gravity point of view, a black hole is special in that an appropriate treatment of the nonperturbative gauge redundancies is vital in obtaining the correct physics (although a similar situation may also occur in cosmology~\cite{Nomura:2011dt,Bousso:2011up}).
It is, therefore, particularly important in the black hole physics not to conflate pictures deriving from different gauge fixings.
We thus begin our discussion with short descriptions of the two constructions based on two different ways of fixings the nonperturbative gauge redundancies.

\subsection*{``Global gauge'' construction}

This construction starts from the conventional global spacetime picture of general relativity.
Perturbative quantization may begin with equal-time hypersurfaces that extend smoothly to both the exterior and interior of the black hole~\cite{Lowe:1995ac,Giddings:2006sj}.
In this picture, the black hole manifestly has the interior region, as implied by general relativity.

The problem, however, is to see its compatibility with unitarity~\cite{Hawking:1976ra,Mathur:2009hf,Almheiri:2012rt}.
First, explicit semiclassical calculation seems to indicate that radiation emitted from the black hole is maximally entangled with modes in the interior, violating the unitarity of the S-matrix~\cite{Hawking:1976ra}.
Second, even if we postulate that Hawking radiation somehow carries information about collapsing/infalling matter to save unitarity, it then leads to the problem of cloning:\ the information about fallen matter is duplicated to Hawking radiation, violating the no-cloning theorem of quantum mechanics~\cite{Wootters:1982zz}.
Finally, depending on hypersurfaces one chooses, the interior of the black hole can be viewed as having an ever increasing spatial volume~\cite{Christodoulou:2014yia,Christodoulou:2016tuu}.
The existence of such large space does not seem to be consistent with the Bekenstein-Hawking entropy~\cite{Bekenstein:1973ur,Hawking:1974sw}, bounding the number of independent black hole states by the horizon area.

Recently, there has been significant progress in addressing these issues.
In particular, new saddles in the gravitational path integral, called replica wormholes, were discovered which contribute to the calculation of entropies~\cite{Penington:2019kki,Almheiri:2019qdq}.
Because of this contribution, naively orthogonal black hole microstates $\ket{\psi_I}$ ($I = 1,\cdots,{\cal K}$) develop overlaps of the form
\begin{equation}
  |\inner{\psi_I}{\psi_J}|^2 = \delta_{IJ} + O\bigl(e^{-S_0}\bigr),
\label{eq:path-overlap}
\end{equation}
where $S_0$ is the coarse-grained entropy of the system, in this case the black hole.
While these overlaps are exponentially suppressed, for ${\cal K} > e^{S_0}$ they add up and after appropriate ``diagonalization'' lead only to $e^{S_0}$ independent states; all the other states are either null or not independent.
Due to this phenomenon, seemingly independent interior states are actually not independent, and as a result the von~Neumann entropy of Hawking radiation follows~\cite{Penington:2019npb,Almheiri:2019psf,Almheiri:2019hni} the Page curve~\cite{Page:1993wv}.
It is plausible that a part of this reduction of state space is associated with the nonperturbative gravitational gauge redundancies~\cite{Marolf:2020xie,Jafferis:2017tiu,McNamara:2020uza}, although a part of them may not.
This reduction also implies that for an old black hole, the degrees of freedom in the interior are not independent of those in the exterior, as anticipated earlier~\cite{Maldacena:2013xja}.

A puzzling feature of this analysis is that the path integral gives
\begin{equation}
  \inner{\psi_I}{\psi_J} = \delta_{IJ},
\label{eq:path-ortho}
\end{equation}
which is inconsistent with Eq.~(\ref{eq:path-overlap}) if both are taken at face value.
Reference~\cite{Penington:2019kki} interpreted this to mean that the gravitational path integral actually computes the average of a quantity over some ensemble, which makes the two equations compatible.
This, however, brings the question:\ what ensemble does the path integral represent?
In this paper, we will discuss how the relevant ensemble can emerge in a consistent unitary theory of gravity.

Summarizing, the basic idea of the construction is to start from a description that is highly redundant but is based on familiar global spacetime of general relativity.
The price to pay is that in order to see truly independent degrees of freedom---and hence unitarity of the theory---one must take into account huge nonperturbative gravitational gauge redundancies, including those relating different spatial topologies~\cite{Marolf:2020xie,Jafferis:2017tiu,McNamara:2020uza}.
If this is done carefully, however, one can see the correct physics as demonstrated recently~\cite{Penington:2019npb,Almheiri:2019psf,Almheiri:2019hni,Penington:2019kki,Almheiri:2019qdq}.

\subsection*{``Unitary gauge'' construction}

An alternative construction is to start from a manifestly unitary description~\cite{Page:1993wv,tHooft:1990fkf,Susskind:1993if}.
This naturally arises in a ``distant description'' of a black hole.
When viewed from a distance, the black hole carries Hawking cloud around it, whose intrinsic energy scale (local, or Tolman~\cite{Tolman:1930zza,Tolman:1930ona}, temperature) increases toward the horizon because of large gravitational blueshift.
This scale reaches the string scale on a surface a microscopic distance away from the classical horizon.
This surface is called the stretched horizon~\cite{Susskind:1993if}, at which the semiclassical description of spacetime breaks down.
Since the dynamics around the stretched horizon cannot be described by a low energy theory, one may consistently assume that the physics there, which is responsible for the Hawking phenomenon, is unitary.

A virtue of this picture is that it is intuitive.
The evolution is unitary by construction (or one can say, as suggested by the AdS/CFT correspondence~\cite{Maldacena:1997re,Gubser:1998bc,Witten:1998qj}), and the interpretation of the Bekenstein-Hawking entropy is simple:\ it basically comes from qubits of a Planck density on the stretched horizon.
Most of the gravitational gauge redundancies are regarded as being fixed, leaving only the standard diffeomorphism and perhaps a little more.

The issue in this picture is to understand how the interior spacetime of a black hole, whose existence is implied by the equivalence principle, may emerge~\cite{Almheiri:2012rt,Almheiri:2013hfa,Marolf:2013dba}.
In this description, the stretched horizon behaves---in a sense---as a surface of regular material such as a piece of coal.
While this makes unitarity manifest, the existence of interior spacetime is obscured; we know that the surface of a piece of coal does not allow for an object to fall freely into it.
What is special about the stretched horizon, making it distinct from a regular material surface?

This problem was addressed in Refs.~\cite{Nomura:2018kia,Nomura:2019qps,Nomura:2019dlz}.
There are two important characteristics that emerge when a black hole is formed.
First, because of large gravitational redshift, energy gaps between different black hole microstates become exponentially small, as measured in the asymptotic region.
This means that the density of states becomes exponentially large~\cite{Bekenstein:1973ur,Hawking:1974sw}.
Second, the intrinsic energy scale (the local temperature of Hawking cloud) becomes large, of order the string scale, near the stretched horizon.
The dynamics at such energies is believed to be maximally quantum chaotic~\cite{Maldacena:2015waa}, fast scrambling~\cite{Hayden:2007cs,Sekino:2008he}, and not to have a feature discriminating low energy species, such as a global symmetry~\cite{Banks:2010zn,Harlow:2018jwu}.
These dynamical features play a crucial role in the emergence of the interior spacetime.

Because of the dynamical properties described above, the state of a black hole quickly becomes a typical state in the relevant microcanonical ensemble, which treats all low energy species in a universal manner.
The prescription in Refs.~\cite{Nomura:2018kia,Nomura:2019qps,Nomura:2019dlz} then allows us to erect an effective theory of the interior on each microstate, which can describe the fate of an object falling into the black hole.
This construction does not work for a regular material surface because of the lack of the typicality and universality, hence singling out the stretched horizon.
An important point is that the construction of the effective theory is restricted to the modes in the near black hole region, called the zone, whose characteristic frequencies $\omega$ are sufficiently, e.g.\ of $O(10)$, larger than the Hawking temperature, $T_{\rm H}$.
This structure makes it possible, unlike the scenarios discussed in Refs.~\cite{Papadodimas:2012aq,Papadodimas:2013jku,Papadodimas:2015jra,Verlinde:2012cy,Verlinde:2013qya,Bousso:2013ifa,Marolf:2015dia}, that the operators describing the interior are state independent, i.e.\ standard linear operators defined throughout the space of microstates, up to exponentially suppressed corrections of order $e^{-\omega/T_{\rm H}}$.
In fact, Ref.~\cite{Nomura:2019dlz} argued that the existence of such globally defined operators is necessary for the emergence of semiclassical physics, which has an intrinsic ambiguity of $O(e^{-\omega/T_{\rm H}})$.

An erection of an effective theory of the interior involves coarse graining:\ many different states in the microscopic theory correspond to a single state in the effective theory.
This coarse graining is the root of the apparent uniqueness of the infalling vacuum, despite the existence of exponentially many black hole microstates.
In this paper, we will see that this also elucidates the ensemble nature of the gravitational path integral in the global gauge construction.

The effective theory erected over the state of the system at a given time (without invoking boundary time evolution in the language of holography~\cite{tHooft:1993dmi,Susskind:1994vu,Bousso:2002ju}) describes only the causal region associated with the black hole zone at that time~\cite{Nomura:2018kia}, giving a specific realization of the idea of black hole complementarity~\cite{Susskind:1993if,Susskind:1993mu}.
For an old black hole, the erection of the effective theory must involve Hawking radiation emitted earlier, as suggested in Ref.~\cite{Maldacena:2013xja}, although the detailed realization is different from that contemplated there:\ the degrees of freedom describing the second exterior of the effective two-sided geometry must also involve black hole ``soft mode'' degrees of freedom~\cite{Nomura:2018kia,Nomura:2019qps,Nomura:2019dlz,Kim:2020cds}.
The non-local identification of degrees of freedom needed here is understood to come from gauge fixing adopted by the unitary gauge construction.

To analyze the relation between this picture and the results of Refs.~\cite{Penington:2019npb,Almheiri:2019psf,Almheiri:2019hni,Penington:2019kki,Almheiri:2019qdq} adopting entanglement wedge reconstruction~\cite{Czech:2012bh,Wall:2012uf,Headrick:2014cta,Jafferis:2015del,Dong:2016eik,Cotler:2017erl}, we need to take into account the effect of time evolution~\cite{Nomura:2019dlz}.
As we will see in this paper, this reveals a structure of bulk reconstruction that is not manifest in the simplest entanglement wedge consideration:\ the amount of information one can reconstruct from radiation at a given time depends on the spacetime location within the entanglement wedge.
This feature originates from the fact that the reconstruction employs a quantum information theoretic protocol~\cite{Hayden:2007cs} which, unlike the erection of an effective interior theory, uses boundary time evolution.
We elucidate this relation in detail and discuss properties of the entanglement wedge reconstruction that are contrasted with those of the effective interior theory, which uses the black hole soft modes in addition to radiation degrees of freedom.

\subsection*{Outline of the paper}

In this paper, we mainly analyze the unitary gauge construction to understand features of the global gauge construction.
In Section~\ref{sec:review}, we begin with discussion of the unitary gauge construction, reviewing relevant aspects of the description in Refs.~\cite{Nomura:2018kia,Nomura:2019qps,Nomura:2019dlz}.
In section~\ref{sec:ensemble}, we find that this construction exhibits the same ensemble property as that found in Ref.~\cite{Penington:2019kki}, providing an understanding of an origin of the ensemble nature of the gravitational path integral.
In Section~\ref{sec:interior}, we investigate reconstruction of the interior.
We discuss two possible ways to reconstruct the interior---effective theories of the interior and entanglement wedge reconstruction---and study their relation.
Finally, we conclude in Section~\ref{sec:concl}.
The two appendices contain details of the stretched horizon, soft modes, and the analysis of signal propagation and scrambling times.

Throughout the paper, we focus on a spherically symmetric, non-near extremal black hole in asymptotically flat or AdS spacetime.%
\footnote{We, however, expect that our basic argument applies to a near extremal black hole as well.
 We also expect that a similar analysis applies to the cosmological horizon of de~Sitter spacetime, where the global gauge and unitary gauge constructions correspond to the treatments in Refs.~\cite{Chen:2020tes,Hartman:2020khs} and \cite{Nomura:2019qps,Nomura:2011dt}, respectively.}
We adopt natural units $c = \hbar = 1$, and we use $l_{\rm s}$ to denote the string length.

\vspace{4mm}

{\bf Note added:} While completing this paper, Ref.~\cite{Liu:2020jsv} appeared which has some overlap with Section~\ref{sec:ensemble} of this paper.

\section{Unitary Gauge Construction}
\label{sec:review}

In this section, we discuss the unitary gauge construction.
This is mostly a review of the description of an evaporating black hole in Refs.~\cite{Nomura:2018kia,Nomura:2019qps,Nomura:2019dlz}, with the focus on aspects relevant to our discussion.
The description has several important elements.
First, it is based on the unitary gauge construction.
Namely, the system is described from a distant perspective, and the degrees of freedom outside the stretched horizon comprise the entire system~\cite{Susskind:1993if}.%
\footnote{In the distant description, the distribution of the degrees of freedom associated with the black hole (soft modes) is peaked toward the stretched horizon, so one may say ``on and outside the stretched horizon'' instead of ``outside the stretched horizon.''
 In this paper, we adopt the latter for brevity.}
Here, the stretched horizon is defined as the surface at which the blueshifted, local Hawking temperature becomes the string scale; see Appendix~\ref{app:sh-sm} for details.
This implies that the quantum mechanical evolution is unitary among the degrees of freedom outside the stretched horizon.
In the context of holography, this corresponds to a description based on the boundary Hamiltonian,%
\footnote{The boundary here need not be the conformal boundary of AdS or an asymptotic infinity in flat spacetime, but can be the holographic screen in Refs.~\cite{Nomura:2016ikr,Nomura:2018kji,Murdia:2020iac}, although in the latter case a fully unitary description may require an ad~hoc introduction of extra ingredients, such as a superposition of different geometries.}
which we may regard as the ``fundamental'' description.

Second, modes of a low energy quantum field in the black hole zone region ($r_{\rm s} \leq r \leq r_{\rm z}$ in Appendix~\ref{app:sh-sm}) are decomposed into hard and soft modes.
The hard modes have frequencies $\omega$ and gaps among them $\varDelta\omega$ sufficiently, e.g.\ a factor of $O(10)$, larger than the Hawking temperature $T_{\rm H}$ (as measured at the same location), while the soft modes have smaller frequencies.
Semiclassical theory is supposed to describe the physics of the hard modes, while the soft modes represent the black hole microstates associated with the semiclassical (black hole) vacuum.
This separation of modes is not quite necessary if we are only interested in basic consequences of unitarity, e.g.\ the Page curve, since for this purpose an excitation of the hard modes can be safely ignored.
It is, however, crucial if we want to discuss the emergence of effective semiclassical spacetime describing the black hole interior~\cite{Nomura:2019dlz} as we now see.

\subsection{Semiclassical spacetime from approximate state independence}
\label{subsec:state-indep}

Consider the space ${\cal H}_M$ of pure states in which the energy $E$ in a spatial region is bounded by $E < M$, where $M$ is sufficiently large that a typical state in ${\cal H}_M$ is a black hole vacuum state.
We then consider the space of all states that are obtained by acting appropriately smoothed hard mode (semiclassical) operators on any of the microstates in ${\cal H}_M$ and have energies smaller than $M + \delta E$.
This space, denoted by $B_{\delta E} {\cal H}_M$, has dimension $e^{S_{\rm exc}}\, {\rm dim}{\cal H}_M$, where $S_{\rm exc}$ is the entropy of the possible semiclassical excitations.
One can then show~\cite{Marolf:2015dia} that a typical state $\ket{\psi}$ in ${\cal H}_{M + \delta E}$ can be written as
\begin{equation}
  \ket{\psi} = \sin\theta\, \ket{\psi_{\rm exc}} + \cos\theta\, \ket{\psi_{\rm vac}}
\label{eq:decomp}
\end{equation}
with
\begin{equation}
  \sin^2\!\theta \sim e^{-\left( \frac{\delta E}{T_{\rm H}} - S_{\rm exc} \right)}.
\label{eq:atypical}
\end{equation}
Here, $\ket{\psi_{\rm exc}}$ and $\ket{\psi_{\rm vac}}$ are elements of $B_{\delta E} {\cal H}_M$ and its complement ${\cal H}_{M + \delta E} / B_{\delta E} {\cal H}_M$, respectively.
Assuming that semiclassical excitations are well within the Bekenstein bound~\cite{Bekenstein:1980jp,Casini:2008cr}, i.e.\ $S_{\rm exc} < \delta E/T_{\rm H}$ with $(\delta E/T_{\rm H} - S_{\rm exc})/S_{\rm exc} \nll 1$, and that a semiclassical excitation has entropy of order a few or larger, we obtain
\begin{equation}
  \mbox{a few} \lesssim S_{\rm exc} < \frac{\delta E}{T_{\rm H}} 
\quad\Rightarrow\quad
  \sin^2\!\theta \ll 1.
\label{eq:semicl}
\end{equation}
We thus find that a state having excitations over a semiclassical black hole background is atypical in the microscopic Hilbert space.

The direct application of the above analysis is limited to the excitations outside the horizon, which raise the energy of the state as measured in the asymptotic region.
However, the conclusion that a semiclassically excited state is microscopically atypical also persists for the states describing the interior obtained by the prescription in Refs.~\cite{Nomura:2018kia,Nomura:2019qps,Nomura:2019dlz}.
This is because such states are produced by ``evolving'' the states described above by an operator that is approximately unitary over the relevant timescale.
This operator does not commute with the boundary Hamiltonian and induces components whose energies measured in the asymptotic region are lower than that of the corresponding vacuum state.
However, the evolution still brings the original vacuum states to the vacuum states in the interior, and so for the excited states as well, preserving the atypicality of the excited states at the microscopic level.

This atypicality of excited states is in stark contrast with the claim in Ref.~\cite{Marolf:2015dia}, which is derived from the picture in Refs.~\cite{Papadodimas:2012aq,Papadodimas:2013jku,Papadodimas:2015jra}.
In particular, unlike the scenarios considered in Refs.~\cite{Papadodimas:2012aq,Papadodimas:2013jku,Papadodimas:2015jra,Verlinde:2012cy,Verlinde:2013qya,Bousso:2013ifa,Marolf:2015dia}, the structure described above implies that the Hilbert space for semiclassical excitations, ${\cal H}_{\rm exc}$, built on each of the orthogonal microstates of the black hole and radiation emitted from it need not overlap significantly with each other.
Indeed, from genericity consideration, we expect that states representing the same semiclassical excitation but built on different orthogonal microstates $A$ and $B$ have overlap
\begin{equation}
  {}_A \bra{\Psi(M)} {\cal O}_{\delta E}^{(A)\dagger} {\cal O}_{\delta E}^{(B)} \ket{\Psi(M)}_B 
  \approx O\biggl( \frac{1}{\sqrt{e^{S_{\rm bh}(M) + S_{\rm rad}}}} e^{-\frac{\delta E}{2 T_{\rm H}}} \biggr),
\label{eq:overlap}
\end{equation}
where $\ket{\Psi(M)}_{A,B}$ represent the microstates of the black hole and radiation, on which semiclassical theories are built~\cite{Nomura:2018kia,Nomura:2019qps,Nomura:2019dlz}, and $S_{\rm bh}(M)$ and $S_{\rm rad}$ are the coarse-grained entropies of the black hole and radiation, respectively.
${\cal O}_{\delta E}^{(A,B)}$ are the suitably normalized operators that excite an appropriately smoothed semiclassical mode of energy $\delta E$ in the black hole region.

\begin{figure}[t]
\begin{center}
  \includegraphics[width=14cm]{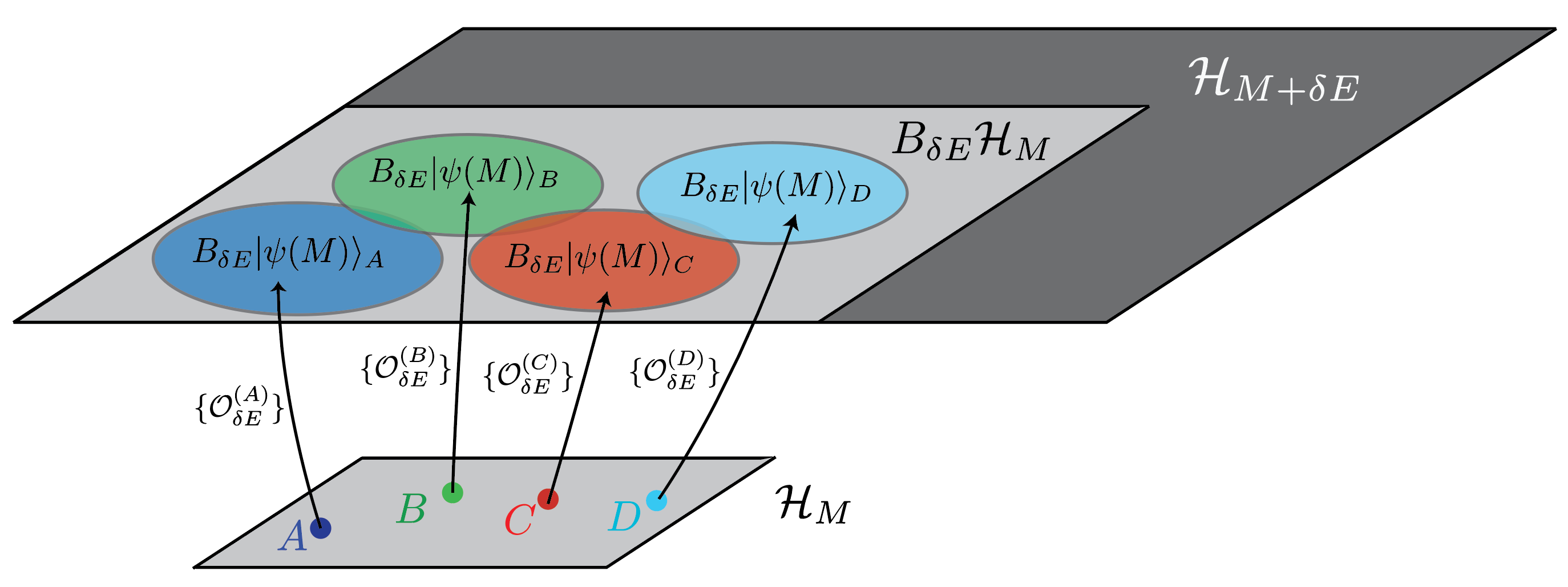}
\end{center}
\caption{The Fock spaces built on microstates of a black hole of mass $M$, which dominate the microcanonical ensemble ${\cal H}_M$, do not significantly overlap with each other because of the energy restriction imposed on hard modes.
 Furthermore, the collection of excited states, $B_{\delta E} {\cal H}_M$, forms only an exponentially small subset of the microcanonical ensemble, ${\cal H}_{M + \delta E}$, of the same energy.}
\label{fig:Hilbert}
\end{figure}

While the overlap in Eq.~(\ref{eq:overlap}) is suppressed by the large exponential factor $e^{-(S_{\rm bh}(M)+S_{\rm rad})/2}$, this by itself is not enough to suppress the overlap between Fock spaces built on different microstates.
To see this, we can consider the probability for the state ${\cal O}_{\delta E}^{(A)} \ket{\Psi(M)}_A$ to overlap with the corresponding semiclassical state built on some other microstate orthogonal to $\ket{\Psi(M)}_A$:%
\footnote{Recall that $\ket{\Psi(M)}_{A,B}$ represent microstates of the black hole {\it and} radiation with the black hole put in the semiclassical vacuum, so that a generic state in the Hilbert space of dimension $e^{S_{\rm bh}(M) + S_{\rm rad}}$ has the black hole of mass $M$.
Note that since black hole evaporation is a thermodynamically irreversible process~\cite{Zurek:1982zz,Page:1983ug}, most of these microstates do not become a state with a larger black hole in empty space when evolved backward in time---there is always junk radiation around it.
This, however, does not change the fact that there are $e^{S_{\rm bh}(M) + S_{\rm rad}}$ independent microstates relevant for the discussion here.}
\begin{equation}
    \sum_{B = 1, B \neq A}^{e^{S_{\rm bh}(M) + S_{\rm rad}}} \left| {}_A \bra{\Psi(M)} {\cal O}_{\delta E}^{(A)\dagger} {\cal O}_{\delta E}^{(B)}\ket{\Psi(M)}_B \right|^2 \approx O\Bigl( e^{-\frac{\delta E}{T_{\rm H}}} \Bigr).
\end{equation}
We find that the exponential factor associated with the microstate entropies indeed disappears.
However, the exponential factor $e^{-\delta E/T_{\rm H}}$ associated with the energy of the excitation remains.
This implies that Fock spaces of hard modes built on different orthogonal microstates are orthogonal up to corrections exponentially suppressed in $\delta E/T_{\rm H}$.
This allows us to treat the microscopic Hilbert space as
\begin{equation}
  {\cal H} \approx {\cal H}_{\rm exc} \otimes {\cal H}_{\rm vac},
\label{eq:prod}
\end{equation}
where the elements of ${\cal H}_{\rm vac}$ cannot be discriminated as quantum degrees of freedom in semiclassical theory.
In other words, semiclassical theory is the theory describing the physics associated with the 
${\cal H}_{\rm exc}$ factor, which is insensitive to the microscopic 
physics occurring in the ${\cal H}_{\rm vac}$ part.
The structure of Hilbert space described here is depicted in Fig.~\ref{fig:Hilbert}.

The structure of the Hilbert space given above makes it possible that for an operator ${\cal O}_{IJ}$ in the semiclassical theory, we can define global operator $\hat{\cal O}_{IJ}$ in the microscopic Hilbert space which acts linearly throughout the space of all microstates:
\begin{equation}
  \hat{\cal O}_{IJ} = \sum_{A = 1}^{e^{S_{\rm bh}(M) + S_{\rm rad}}} 
    {\cal O}_{IJ}\, \ket{\Psi_I(M)}_A\, {}_A \bra{\Psi_J(M)},
\label{eq:linear-op}
\end{equation}
where $A$ runs over orthogonal vacuum microstates, $I$ and $J$ are the indices specifying semiclassical states (regardless of the microstate), and $\ket{\Psi_I(M)}_A$ is the semiclassical state $I$ built on microstate $A$.
In Ref.~\cite{Nomura:2019dlz}, it was conjectured that the emergence of semiclassical physics requires the existence of these approximately state-independent operators.%
\footnote{This can be viewed as an extension of an analogous leading-order statement about the geometry~\cite{Almheiri:2016blp,Nomura:2017fyh} to higher order in the Newton constant expansion, including matter excitations.}
These operators obey the algebra of the semiclassical theory throughout the microstates, up to corrections of order $e^{-E_{\rm exc}/T_{\rm H}}$ where $E_{\rm exc}$ is the energy of the semiclassical excitation.
These corrections are the intrinsic ambiguity of the semiclassical theory, requiring $E_{\rm exc}$ to be sufficiently larger than $T_{\rm H}$.

\subsection{Black hole interior from chaotic dynamics at the horizon}
\label{subsec:chaotic}

Another important ingredient of the description in Refs.~\cite{Nomura:2018kia,Nomura:2019qps,Nomura:2019dlz} is the proposed UV-IR relation, in particular the role chaotic dynamics at the horizon plays for the emergence of interior spacetime.

It is widely believed that the dynamics of the stretched horizon in a distant picture is maximally quantum chaotic~\cite{Maldacena:2015waa}.
It is also expected that this dynamics does not respect any global symmetry~\cite{Banks:2010zn,Harlow:2018jwu}.
The claim of Refs.~\cite{Nomura:2018kia,Nomura:2019qps,Nomura:2019dlz} is that these properties of the stretched horizon dynamics are crucial for the emergence of interior spacetime.
In fact, together with the exponential degeneracy of states, the appearance of such dynamics can be viewed as the quantum mechanical analog of gravitational collapse in classical general relativity.

To see how this works, consider a system with a black hole, whose mass $M$ is determined by the maximal precision of
\begin{equation}
  \Delta \approx O(2\pi T_{\rm H})
\end{equation}
allowed by the uncertainty principle.
At a given time $t$, the state of the system---with the black hole being put in the semiclassical vacuum---is given by
\begin{equation}
  \ket{\Psi(M)} = \sum_n \sum_{i_n = 1}^{e^{S_{\rm bh}(M-E_n)}} 
    \sum_{a = 1}^{e^{S_{\rm rad}}} c_{n i_n a} \ket{\{ n_\alpha \}} 
    \ket{\psi^{(n)}_{i_n}} \ket{\phi_a},
\label{eq:sys-state}
\end{equation}
where the state is assumed to be normalized
\begin{equation}
  \sum_n \sum_{i_n = 1}^{e^{S_{\rm bh}(M-E_n)}} \sum_{a = 1}^{e^{S_{\rm rad}}} 
    |c_{n i_n a}|^2 = 1.
\label{eq:norm}
\end{equation}
The three factors in the right-hand side of Eq.~(\ref{eq:sys-state}) represent states of the hard modes, soft modes, and modes outside the zone region (far modes), respectively.
The notations and basic properties for these states are the following:%
\footnote{By construction, the system of a black hole and radiation evolves unitarily with the state taking the form of Eq.~(\ref{eq:sys-state}) at each moment in time.
 In particular, the entanglement entropy between the black 
hole and radiation $S^{\rm vN}_{\rm hard + soft} = S^{\rm vN}_{\rm rad}$ follows the Page curve~\cite{Page:1993wv}, where $S^{\rm vN}_A$ is the von~Neumann entropy of subsystem $A$.}
\begin{itemize}
\item
$\ket{\{ n_\alpha \}}$ are orthonormal states of the hard modes, with $n \equiv \{ n_\alpha \}$ representing the set of all occupation numbers $n_\alpha$ ($\geq 0$).
The index $\alpha$ collectively denotes the species, frequency, and angular-momentum quantum numbers of a mode, and $E_n$ is the energy of the state $\ket{\{ n_\alpha \}}$ as measured in the asymptotic region (with precision $\Delta$).
\item
$\ket{\psi^{(n)}_{i_n}}$ are orthonormal states of the soft modes entangled with $\ket{\{ n_\alpha \}}$.
Because of the energy constraint imposed on the black hole, these states have energies $M-E_n$, with precision $\Delta$.
Assuming that the density of hard mode states is negligible compared with that of the soft modes (which is justified as we are only interested in states that do not yield significant backreaction),%
\footnote{In fact, we expect that the logarithm of the dimension of the hard mode Hilbert space, $\ln {\rm dim}{\cal H}_{\rm hard}$, is much smaller than $S_{\rm bh}(M-E_n)$ and $S_{\rm rad}$ throughout the history of the black hole.}
the density of soft mode states is given by the standard Bekenstein-Hawking formula, implying that $i_n$ runs over the $n$-dependent range
\begin{equation}
  i_n = 1,\cdots,e^{S_{\rm bh}(M-E_n)}.
\label{eq:i_n}
\end{equation}
With the black hole put in the semiclassical vacuum, any extra attribute a hard mode state may have, e.g.\ a charge or angular momentum, is compensated by that of the corresponding soft mode states (within the precision allowed by the uncertainty principle).
This implies that soft mode states associated with different hard mode states are virtually orthogonal:
\begin{equation}
  \inner{\psi^{(m)}_{i_m}}{\psi^{(n)}_{j_n}} = \delta_{m n} \delta_{i_m j_n}.
\label{eq:soft-orthonorm}
\end{equation}
\item
$\ket{\phi_a}$ represents the set of orthonormal states representing the system in the far region $r > r_{\rm z}$.
We assume that these are fully specified by the states of Hawking radiation emitted earlier, i.e.\ emitted from $r \approx r_{\rm z}$ to the asymptotic region before time $t$, although this assumption is not essential.
$S_{\rm rad}$ in Eq.~(\ref{eq:sys-state}) is then the coarse-grained entropy of the early radiation.
\end{itemize}

The spatial distribution of the soft modes is determined by the blueshifted local Hawking temperature; see Appendix~\ref{app:sh-sm}.
It is peaked toward the stretched horizon, at which the local temperature is of order the string scale and the dynamics is maximally chaotic.%
\footnote{Recall that this dynamics cannot be described by a low energy theory.
 In fact, the internal dynamics near the stretched horizon is expected to be nonlocal in the spatial directions along the horizon~\cite{Hayden:2007cs,Sekino:2008he}.
 This nonlocality arises presumably from gauge fixing performed to go to the unitary gauge description.}
We thus expect that the coefficients $c_{n i_n a}$ in Eq.~(\ref{eq:sys-state}) take generic values in the spaces of the hard and soft modes.
In particular, statistically on average
\begin{equation}
  |c_{n i_n a}| \sim \frac{1}{\sqrt{S_{\rm tot}}},
\label{eq:c-gen}
\end{equation}
where
\begin{equation}
  S_{\rm tot} 
  \equiv \biggl( \sum_n e^{S_{\rm bh}(M-E_n)} \biggr)\, e^{S_{\rm rad}} 
  = \biggl( \sum_n e^{-\frac{E_n}{T_{\rm H}}} \biggr)\, 
    e^{S_{\rm bh}(M)} e^{S_{\rm rad}}.
\label{eq:S_tot}
\end{equation}
With $c_{n i_n a}$ being complex numbers, their phases are uniformly distributed and the variance of $c_{n i_n a}$ is comparable to the average of $|c_{n i_n a}|$.
Furthermore, because of the fast scrambling nature of the dynamics~\cite{Hayden:2007cs,Sekino:2008he}, these configurations are reached quickly.
The standard thermal nature of the black hole is then obtained upon tracing out the soft modes:
\begin{equation}
  {\rm Tr}_{\rm soft} \ket{\Psi(M)} \bra{\Psi(M)}
  = \frac{1}{\sum_m e^{-\frac{E_m}{T_{\rm H}}}} \sum_n 
    e^{-\frac{E_n}{T_{\rm H}}} \ket{\{ n_\alpha \}} \bra{\{ n_\alpha \}} 
    \otimes \rho_{\phi,n},
\label{eq:rho_HR}
\end{equation}
where $\rho_{\phi,n}$ are reduced density matrices for the early radiation, whose $n$-dependence is small and of order $1/\sqrt{e^{S_{\rm bh}(M)}}$. 
Note that in order to obtain the correct Boltzmann factor, $\propto e^{-E_n/T_{\rm H}}$, it is essential that the coefficients $c_{n i_n a}$ take generic values across all low energy species, i.e.\ for $n$ to run over all low energy species~\cite{Nomura:2019qps}.%
\footnote{Here, low energy species mean those below the local Hawking temperature.
 In particular, near the stretched horizon they include all species below the string scale $1/l_{\rm s}$.}

The form of the state in Eq.~(\ref{eq:sys-state}) with the statistic properties described above allows for the following coarse-grained description of the dynamics of the hard modes.
Given the state of the system at time $t$, Eq.~(\ref{eq:sys-state}), we can define a set of coarse-grained states each of which is entangled with a specific hard mode state:
\begin{equation}
  \ketc{\{ n_\alpha \}} 
  \propto \sum_{i_n = 1}^{e^{S_{\rm bh}(M-E_n)}} \sum_{a = 1}^{e^{S_{\rm rad}}} 
    c_{n i_n a} \ket{\psi^{(n)}_{i_n}} \ket{\phi_a},
\label{eq:coarse-prop}
\end{equation}
where we have used the same label as the corresponding hard mode state to specify the coarse-grained state, which we denote by the double ket symbol.
Using Eq.~(\ref{eq:c-gen}), the squared norm of the (non-normalized) state on the right-hand side of Eq.~(\ref{eq:coarse-prop}) is given by
\begin{equation}
  \sum_{i_n = 1}^{e^{S_{\rm bh}(M-E_n)}} 
    \sum_{a = 1}^{e^{S_{\rm rad}}} |c_{n i_n a}|^2 
 = \frac{e^{-\frac{E_n}{T_{\rm H}}}}
    {\Bigl( \sum_m e^{-\frac{E_m}{T_{\rm H}}} \Bigr)} 
    \left[ 1 + O\biggl( \frac{1}{\sqrt{e^{S_{\rm bh}(M-E_n)} e^{S_{\rm rad}}}} 
      \biggr) \right]
\label{eq:coarse-norm}
\end{equation}
for generic black hole and radiation microstates.
Here, the second term in the square brackets represents the size of statistical fluctuations over different microstates.
The normalized coarse-grained state $\ketc{\{ n_\alpha \}}$ is thus given by
\begin{equation}
  \ketc{\{ n_\alpha \}} 
  = e^{\frac{E_n}{2 T_{\rm H}}} \sqrt{\sum_m e^{-\frac{E_m}{T_{\rm H}}}}\, 
    \sum_{i_n = 1}^{e^{S_{\rm bh}(M-E_n)}} \sum_{a = 1}^{e^{S_{\rm rad}}} 
    c_{n i_n a} \ket{\psi^{(n)}_{i_n}} \ket{\phi_a},
\label{eq:coarse}
\end{equation}
up to a fractional correction of order $e^{-(S_{\rm bh}(M-E_n) + S_{\rm rad})/2} \sim e^{-S_{\rm tot}/2}$ in the overall normalization.

With this coarse graining, the state of the system in Eq.~(\ref{eq:sys-state}) can be written as
\begin{equation}
  \ketc{\Psi(M)} 
  = \frac{1}{\sqrt{\sum_m e^{-\frac{E_m}{T_{\rm H}}}}} 
    \sum_n e^{-\frac{E_n}{2 T_{\rm H}}} 
    \ket{\{ n_\alpha \}} \ketc{\{ n_\alpha \}}
\label{eq:BH-coarse}
\end{equation}
{\it regardless of the values of $c_{n i_n a}$}, which takes the form of the standard thermofield double state of the two-sided black hole~\cite{Unruh:1976db,Israel:1976ur}.
We can therefore build the effective theory describing the interior on this state, as described in Refs.~\cite{Nomura:2018kia,Nomura:2019qps,Nomura:2019dlz} and will be discussed further in Section~\ref{sec:interior}.
Note that in order to obtain the correct Boltzmann-weight coefficients, $\propto e^{-E_n/2 T_{\rm H}}$, it is crucial that the hard and soft modes are well scrambled, so that $c_{n i_n a}$ take values statistically independent of $n$.
This construction, therefore, works only for a black hole stretched horizon, which does not have a low energy structure.
A regular material surface does not admit an analogous construction because of the lack of this universality and hence does not have near empty interior spacetime.
The coarse graining described here is the origin of the apparent uniqueness of the infalling vacuum, despite the existence of exponentially many black hole microstates.

We stress that the meaning of the coarse graining here is that a single state $\ketc{\{ n_\alpha \}}$ in the semiclassical, effective theory---the left-hand side of Eq.~(\ref{eq:coarse})---corresponds to multiple different microstates---the right-hand side of Eq.~(\ref{eq:coarse})---that depend on the state of the black hole and radiation represented by the coefficients $c_{n i_n a}$.
This introduces a statistical nature in the description of the interior spacetime, despite the fact that the microscopic theory is a single unitary theory that does not have any ensemble aspect at the fundamental level.

\section{Ensemble from Coarse Graining}
\label{sec:ensemble}

In this section, we will see how the unitary gauge construction of the previous section can elucidate the origin of the ensemble nature of the gravitational path integral encountered in the global gauge construction.
In discussing this, we will ignore the existence of hard modes and the energy constraint imposed on the hard and soft mode system, mainly because the analysis of Ref.~\cite{Penington:2019kki}, which we will discuss here, does not consider these elements.

As we have seen in the previous section, the separation between the hard and soft modes as well as the energy constraint imposed on them are vital in understanding the emergence of the interior.
These are, however, not critical in seeing physics associated with unitarity (and hence one could get the correct Page curve without taking these elements into account, as was the case in recent studies).
One way to see this is to perform the Schmidt decomposition in the space of soft-mode and radiation states for each $n$ in the state of Eq.~(\ref{eq:sys-state}):
\begin{equation}
  \ket{\Psi(M)} = \sum_n \sum_{i_n=1}^{{\cal N}_n} 
    c^n_{i_n} \ket{H_n} \ket{S_{n,i_n}} \ket{R_{n,i_n}},
\label{eq:ent-str}
\end{equation}
where $\ket{H_n}$, $\ket{S_{n,i_n}}$, and $\ket{R_{n,i_n}}$ represent the states of the hard modes, soft modes, and radiation, respectively, and
\begin{equation}
  {\cal N}_n = {\rm min}\bigl\{ e^{S_{\rm bh}(M-E_n)}, e^{S_{\rm rad}} \bigr\}.
\label{eq:N_n}
\end{equation}
We see that the entanglement necessary for interior spacetime has to do with the index $n$, while that responsible for unitarity has to do with the summations of indices $i_n$.
In fact, after the Page time (when the unitarity becomes an issue), ${\cal N}_n$ is given by $e^{S_{\rm bh}(M-E_n)}$, so that the entanglement between the black hole and radiation comes mostly from the vacuum index $i_0$ shared between the soft-mode and radiation states.
In particular, hard modes play only a minor role.

This justifies our neglect of hard modes in the discussion in this section.
We will only reinstate them and the energy constraint in the next section, when we discuss reconstruction of the interior.

\subsection{Ensemble in the global gauge construction}
\label{subsec:puzzle}

In Ref.~\cite{Penington:2019kki}, some of the gravitational path integrals are calculated at the nonperturbative level for a simple model of an evaporating black hole.
The authors considered an entangled state between the black hole (soft mode) microstates and early radiation
\begin{equation}
  \ket{\Psi} = \frac{1}{\sqrt{\cal K}} \sum_{I=1}^{\cal K} \ket{\psi_I} \ket{I}
\end{equation}
and computed the trace of powers of the reduced density matrix $\rho_{\rm R}$ of the radiation, which is the same as that of the reduced density matrix $\rho$ of the black hole.
Here, $\ket{\psi_I}$ and $\ket{I}$ ($I = 1,\cdots,{\cal K}$) are an a priori complete set of black hole microstates and the states of the radiation entangled with them.
The result they obtained is
\begin{equation}
  {\rm Tr}(\rho^N) \approx \begin{cases}
    \frac{1}{{\cal K}^{N-1}} & \mbox{for } {\cal K} \ll e^{S_{\rm bh}} \\
     \frac{1}{e^{(N-1)S_{\rm bh}}} & \mbox{for } {\cal K} \gg e^{S_{\rm bh}},
  \end{cases}
\label{eq:Tr_rho-N}
\end{equation}
where $S_{\rm bh}$ is the coarse-grained entropy of the black hole.

A puzzling feature of this result is that while the ${\cal K}$ black hole microstates satisfy
\begin{equation}
  \inner{\psi_I}{\psi_J} = \delta_{IJ}
\label{eq:contra-1}
\end{equation}
by construction, Eq.~(\ref{eq:Tr_rho-N}) implies
\begin{equation}
  |\inner{\psi_I}{\psi_J}|^2 = \delta_{IJ} + O\bigl(e^{-S_{\rm bh}}\bigr)
\label{eq:contra-2}
\end{equation}
because
\begin{equation}
  {\rm Tr}(\rho^2) = \frac{1}{{\cal K}^2} \sum_{I,J=1}^{\cal K} |\inner{\psi_I}{\psi_J}|^2.
\end{equation}
If taken literally, the two equations (\ref{eq:contra-1}) and (\ref{eq:contra-2}) are incompatible.
Reference~\cite{Penington:2019kki} suggested that this apparent contradiction might arise because the ``true'' quantum amplitude is, in fact, given by
\begin{equation}
  \inner{\psi_I}{\psi_J} = \delta_{IJ} + O\bigl(e^{S_{\rm bh}/2}\bigr) R_{IJ},
\end{equation}
where $R_{IJ}$ is a random variable with mean zero, while the gravitational path integral computes some type of average over $R_{IJ}$ so that
\begin{equation}
  \overline{\inner{\psi_I}{\psi_J}} = \delta_{IJ},
\qquad
  \overline{|\inner{\psi_I}{\psi_J}|^2} = \delta_{IJ} + O\bigl(e^{-S_{\rm bh}}\bigr).
\label{eq:average}
\end{equation}
If Eqs.~(\ref{eq:contra-1},~\ref{eq:contra-2}) are interpreted to mean Eq.~(\ref{eq:average}), then there is no real inconsistency.

The idea that the gravitational path integral may in fact be computing some coarse-grained version of quantities, averaged over some microscopic information, has attracted much attention recently~\cite{Bousso:2019ykv,Pollack:2020gfa,Giddings:2020yes,Afkhami-Jeddi:2020ezh,Maloney:2020nni,Belin:2020hea,Cotler:2020ugk,Perez:2020klz,Maxfield:2020ale,Blommaert:2020seb,Gesteau:2020wrk,Bousso:2020kmy,Stanford:2020,Balasubramanian:2020jhl,Engelhardt:2020qpv,Chen:2020tes}.
But if so, what kind of average is it taking and over what ensemble?
The ($1+1$)-dimensional gravity theory studied in Ref.~\cite{Penington:2019kki} is indeed known to be dual to an ensemble of 1-dimensional quantum theories~\cite{Saad:2019lba,Stanford:2019vob,Saad:2019pqd}, but the concept that the dual of a gravitational theory be an ensemble of theories seems to be at odds with what is known for such dualities in higher dimensions~\cite{McNamara:2020uza}.
Below, we see that the statistical nature of the path integral found in Ref.~\cite{Penington:2019kki} can arise from coarse graining~\cite{Nomura:2018kia,Nomura:2019qps,Nomura:2019dlz} needed to erect an effective theory of the interior.

\subsection{Interpretation in the unitary gauge construction}
\label{subsec:solution}

Let us take the unitary gauge construction.
Consider microstates of the form of Eq.~(\ref{eq:sys-state}), in which the black hole is put in the semiclassical vacuum.
As we have seen in Section~\ref{subsec:chaotic}, coarse-grained states of the soft modes and early radiation corresponding to states in the effective second exterior are given by Eq.~(\ref{eq:coarse-prop}).
For our purpose, it is sufficient to focus on the vacuum microstates, i.e.\ the states in which the hard modes are not excited:\ $\forall \alpha, n_\alpha = 0$.%
\footnote{The argument below, however, also applies to the collection of microstates in which the hard modes take an excited configuration.}
These states are given by
\begin{equation}
  \ketc{\Omega} 
  = \sqrt{z}\, \sum_{i = 1}^{e^{S_{\rm bh}(M)}} \sum_{a = 1}^{e^{S_{\rm rad}}} 
    c_{i a} \ket{\psi_i} \ket{\phi_a},
\end{equation}
where we have denoted $\{ \forall \alpha, n_\alpha = 0 \}$ by $\Omega$ and dropped the specification of $\Omega$ in the index $i$, soft mode states $\ket{\psi_i}$, and coefficients $c_{ia}$; $z = \sum_n e^{-E_n/T_{\rm H}}$ is a factor associated with the normalization.
Different microstates correspond to different values of the coefficients $c_{i a}$.
We introduce the index $A$ to specify the microstate so that
\begin{equation}
  \ket{\Omega}_A 
  = \sqrt{z}\, \sum_{i = 1}^{e^{S_{\rm bh}}} \sum_{a = 1}^{e^{S_{\rm rad}}} c^A_{i a} \ket{\psi_i} \ket{\phi_a}.
\label{eq:vacuum-ens}
\end{equation}
Here and below, we drop the argument $M$ of $S_{\rm bh}$.

We now show that the ensemble of soft mode (black hole) microstates defined by the collection of randomly selected states $\ket{\psi}_I$ ($I = 1,\cdots,{\cal K}$)
\begin{equation}
  \ket{\psi}_I = \sum_{i = 1}^{e^{S_{\rm bh}}} d^I_i \ket{\psi_i},
\qquad
  \sum_{i = 1}^{e^{S_{\rm bh}}} |d^I_i|^2 = 1
\label{eq:soft-ens}
\end{equation}
has precisely the feature in Eq.~(\ref{eq:average}), and the reduced density matrices obtained from the ensemble in Eq.~(\ref{eq:vacuum-ens}) reproduces the result in Eq.~(\ref{eq:Tr_rho-N}).
This implies that, while the unitary gauge construction cannot see possible null states that have already been projected out in going to the construction, the ensemble remaining in it is still able to explain the puzzling feature of the gravitational path integral described in Section~\ref{subsec:puzzle}.
Specifically, the ensemble nature of the gravitational path integral arises from the fact that the exponentially dense spectrum of microstates caused by a large redshift makes the semiclassical path integral unable to resolve these microstates.
In particular, the phenomenon does not require an ensemble of different quantum theories.

We first note that the ensemble in Eq.~(\ref{eq:soft-ens}) practically contains the double exponential number ${\cal N}_{\rm bh,eff}$ of ``independent'' states
\begin{equation}
  {\cal N}_{\rm bh,eff} \,\approx\, e^{e^{S_{\rm bh}}} \,\gg\, e^{S_{\rm bh}}.
\end{equation}
To see this, one can compute an inner product of different microstates
\begin{equation}
  {}_I\inner{\psi}{\psi}_J = \sum_{i = 1}^{e^{S_{\rm bh}}} d^{I*}_i d^J_i.
\label{eq:inner}
\end{equation}
Since we statistically expect that $|d^I_i| \sim 1/\sqrt{e^{S_{\rm bh}}}$ with $d^I_i$ having random phases, it is exponentially suppressed for $I \neq J$
\begin{equation}
  {}_I\inner{\psi}{\psi}_J \approx O\biggl( \frac{1}{\sqrt{e^{S_{\rm bh}}}} \biggr).
\label{eq:inner-supp}
\end{equation}
This is the case even if $\ket{\psi}_I$ and $\ket{\psi}_J$ are not orthogonal, unless $|(d^I_i - d^J_i)/d^I_i| \ll 1$ for the majority of $i$ (which requires a double exponential coincidence).
Moreover, when we average Eq.~(\ref{eq:inner}) over the space of microstates using the Haar measure, we get
\begin{equation}
  \overline{{}_I\inner{\psi}{\psi}_J } \,=\, \int dU\, {}_{U(I)}\inner{\psi}{\psi}_{U(J)} \,=\, \int dU \sum_{i = 1}^{e^{S_{\rm bh}}} d^{U(I)*}_i d^{U(J)}_i \,=\, 0
\label{eq:inner-ave}
\end{equation}
if $I \neq J$ (barring a possible double exponentially suppressed coincidence).
Here, $U(I)$ represents the state obtained by acting a unitary rotation $U$ on the state $I$ in the space of microstates of dimension $e^{S_{\rm bh}}$.
Therefore, if the gravitational path integral is indeed computing the average over the space of microstates, it is meaningful to consider ``independent'' (overcomplete) microstates $\ket{\psi}_I$ ($I = 1,\cdots,{\cal K}$) with ${\cal K} \gg e^{S_{\rm bh}}$, as was done in Ref.~\cite{Penington:2019kki}.

The computation of the second quantity in Eq.~(\ref{eq:average}) goes similarly.
We obtain
\begin{equation}
  \overline{|\inner{\psi_I}{\psi_J}|^2} \,=\, \int dU \sum_{i,j = 1}^{e^{S_{\rm bh}}} d^{U(I)*}_i d^{U(J)}_i d^{U(J)*}_j d^{U(I)}_j \,=\, \delta_{IJ} + O\bigl(e^{-S_{\rm bh}}\bigr).
\end{equation}
We thus find that the feature in Eq.~(\ref{eq:average}) is reproduced by the ensemble of black hole microstates in Eq.~(\ref{eq:soft-ens}).

Let us now consider microstates of the black hole and radiation system given in Eq.~(\ref{eq:vacuum-ens}).
Remember that we are ignoring hard mode excitations, so the states in Eq.~(\ref{eq:vacuum-ens}) with the hard mode vacuum attached (which we omit here) are the appropriate microstates for the system.
For each microstate $A$, the reduced density matrix of the black hole is given by
\begin{equation}
  \rho_A \,=\, {\rm Tr}_{\rm rad} \ket{\Omega}_A {}_A\bra{\Omega} \,=\, z \sum_{i,j = 1}^{e^{S_{\rm bh}}} \sum_{a = 1}^{e^{S_{\rm rad}}} c^A_{i a} c^{A*}_{j a} \ket{\psi_i} \bra{\psi_j}.
\end{equation}
Thus,
\begin{equation}
  {\rm Tr}(\rho_A^N) = z^N \sum_{i_1,\cdots,i_N}^{e^{S_{\rm bh}}} \sum_{a_1,\cdots,a_N}^{e^{S_{\rm rad}}} c^A_{i_N a_1} c^{A*}_{i_1 a_1} c^A_{i_1 a_2} c^{A*}_{i_2 a_2} \,\cdots\,\, c^A_{i_{N-1} a_N} c^{A*}_{i_N a_N}.
\end{equation}
Note the staggered structure for the summations of $i$ and $a$ indices.

To evaluate this, we can consider $\rho_A$ as a matrix in the $(i,j)$ space
\begin{equation}
  (\rho_A)_{ij} \equiv z \sum_{a = 1}^{e^{S_{\rm rad}}} c^A_{i a} c^{A*}_{j a},
\end{equation}
which is an $e^{S_{\rm bh}} \times e^{S_{\rm bh}}$ Hermitian matrix of
\begin{equation}
  {\rm rank}\,(\rho_A)_{ij} = {\rm min}\bigl\{ e^{S_{\rm bh}}, e^{S_{\rm rad}} \bigr\}.
\end{equation}
Using the statistical properties of the coefficients $c^A_{i a}$ discussed around Eqs.~(\ref{eq:c-gen},~\ref{eq:S_tot}), we find that this matrix has the following elements after diagonalization:
\begin{equation}
  (\rho_A)_{ii} \approx O\biggl(\frac{1}{e^{S_{\rm rad}}}\biggr) > 0
\qquad
  (i = 1,\cdots,e^{S_{\rm rad}})
\end{equation}
for $S_{\rm rad} < S_{\rm bh}$, and
\begin{equation}
  (\rho_A)_{ii} = \frac{1}{e^{S_{\rm bh}}} \left( 1 + \delta_i \right)
\qquad
  (i = 1,\cdots,e^{S_{\rm bh}})
\end{equation}
for $S_{\rm rad} > S_{\rm bh}$.
Here,
\begin{equation}
  \delta_i \in \mathbb{R},\quad |\delta_i| \sim \frac{1}{\sqrt{e^{S_{\rm rad}}}},
\end{equation}
and the signs of $\delta_i$ are random.
We thus obtain
\begin{equation}
  \overline{{\rm Tr}(\rho^N)} \,=\, \int dV\, {\rm Tr}\left(\rho_{V(A)}^N\right) \,\approx\, \begin{cases}
    \frac{1}{e^{(N-1)S_{\rm rad}}} & \mbox{for } e^{S_{\rm rad}} \ll e^{S_{\rm bh}} \\
    \frac{1}{e^{(N-1)S_{\rm bh}}} & \mbox{for } e^{S_{\rm rad}} \gg e^{S_{\rm bh}},
  \end{cases}
\label{eq:Tr_rhoN-1}
\end{equation}
where $V$ represents unitaries acting on the space of microstates of dimension $e^{S_{\rm bh}+S_{\rm rad}}$.

To connect this with the analysis of Ref.~\cite{Penington:2019kki}, we note that the soft mode states entangled with different radiation states $\ket{\phi_a}$, i.e.\
\begin{equation}
  \ket{\psi_a}_A \equiv \frac{1}{\sum_{j = 1}^{e^{S_{\rm bh}}} |c^A_{i a}|^2} \sum_{i = 1}^{e^{S_{\rm bh}}} c^A_{i a} \ket{\psi_i} 
\qquad
  (a = 1,\cdots,e^{S_{\rm rad}})
\end{equation}
can all be regarded as independent even if $e^{S_{\rm rad}} > e^{S_{\rm bh}}$, as discussed above (unless $e^{S_{\rm rad}}$ is double exponentially large, $e^{S_{\rm rad}} \gtrsim e^{e^{S_{\rm bh}}}$).
This implies that the number of terms in the maximally entangled state of Ref.~\cite{Penington:2019kki} should be identified as $e^{S_{\rm rad}}$:
\begin{equation}
  {\cal K} \approx e^{S_{\rm rad}}
\end{equation}
regardless of the relative size between $e^{S_{\rm rad}}$ and $e^{S_{\rm bh}}$.
The expression in Eq.~(\ref{eq:Tr_rhoN-1}) can then be written as
\begin{equation}
  \overline{{\rm Tr}(\rho^N)} \approx \begin{cases}
    \frac{1}{{\cal K}^{(N-1)}} & \mbox{for } {\cal K} \ll e^{S_{\rm bh}} \\
    \frac{1}{e^{(N-1)S_{\rm bh}}} & \mbox{for } {\cal K} \gg e^{S_{\rm bh}},
  \end{cases}
\label{eq:Tr_rhoN-2}
\end{equation}
reproducing Eq.~(\ref{eq:Tr_rho-N}).

We emphasize that the result here is obtained in a single unitary theory without invoking an ensemble over different Hamiltonians at the fundamental level.
The semiclassical gravitational path integral simply cannot resolve exponentially degenerate black hole microstates and hence gives results corresponding to the average over these states.
This implies that while the semiclassical gravitational path integral cannot probe details of each microstate, it knows it is dealing with a collection of states.
Indeed, this is consistent with the fact that the Bekenstein-Hawking entropy can be read off using the Euclidean path integral method~\cite{Gibbons:1976ue}.

\section{Reconstructing the Interior}
\label{sec:interior}

In this section, we discuss reconstruction of the interior of an evaporating black hole.
We first review the description of Refs.~\cite{Nomura:2018kia,Nomura:2019qps,Nomura:2019dlz} and see how the black hole interior emerges in the unitary gauge construction without invoking time evolution as viewed from asymptotic infinity.
We then study how this picture is related with entanglement wedge reconstruction, in particular that involving entanglement islands in the global spacetime picture~\cite{Penington:2019npb,Almheiri:2019psf,Almheiri:2019hni,Penington:2019kki,Almheiri:2019qdq}.
We find that reconstruction is not uniform throughout the entanglement wedge.
In particular, the amount of information one can reconstruct from a given radiation state depends on the spacetime position within the entanglement wedge.

\subsection{Effective theory of the interior: reconstruction without time evolution}
\label{subsec:without-ev}

Suppose that the semiclassical vacuum state is given by Eq.~(\ref{eq:sys-state}) at time $t$.
Here, $t$ is the time measured at asymptotic infinity, which we refer to as the boundary time.
With the coarse graining in Eq.~(\ref{eq:coarse}), this state can be written as Eq.~(\ref{eq:BH-coarse}). The question of the interior is if there is a description in which a small excitation falling into the black hole, represented by excitations of hard modes in the zone, sees a smooth horizon and near empty spacetime inside it.

Such a description, if it exists, cannot be the one based on time evolution generated by the boundary Hamiltonian.
In that description, a small object falling into the black hole is absorbed into the stretched horizon once it reaches there, whose information will be later sent back to ambient space by Hawking emission.
To describe the object's experience, we need a different time evolution operator associated with the proper time of the object.

A small object in the zone can be described by standard annihilation and creation operators acting on the hard modes
\begin{align}
  b_\gamma &= \sum_n \sqrt{n_\gamma}\, 
    \ket{\{ n_\alpha - \delta_{\alpha\gamma} \}} \bra{\{ n_\alpha \}},
\label{eq:ann}\\
  b_\gamma^\dagger &= \sum_n \sqrt{n_\gamma + 1}\, 
    \ket{\{ n_\alpha + \delta_{\alpha\gamma} \}} \bra{\{ n_\alpha \}}.
\label{eq:cre}
\end{align}
We thus consider a future evolution of a state obtained by acting appropriately smoothed creation operators $b_\gamma^\dagger$ on a vacuum microstate $\ket{\Psi(M)}$.
We are interested in a time evolution operator that makes the object's experience manifest.

At the coarse-grained level, we can define ``mirror'' operators~\cite{Papadodimas:2012aq}
\begin{align}
  \tilde{b}_\gamma &= \sum_n \sqrt{n_\gamma}\, 
    \ketc{\{ n_\alpha - \delta_{\alpha\gamma} \}} 
    \brac{\{ n_\alpha \}},
\label{eq:ann-m}\\
  \tilde{b}_\gamma^\dagger &= \sum_n \sqrt{n_\gamma + 1}\, 
    \ketc{\{ n_\alpha + \delta_{\alpha\gamma} \}} 
    \brac{\{ n_\alpha \}},
\label{eq:cre-m}
\end{align}
which can be used to form annihilation and creation operators for infalling modes:
\begin{align}
  a_\xi &= \sum_\gamma 
    \bigl( \alpha_{\xi\gamma} b_\gamma 
    + \beta_{\xi\gamma} b_\gamma^\dagger 
    + \zeta_{\xi\gamma} \tilde{b}_\gamma 
    + \eta_{\xi\gamma} \tilde{b}_\gamma^\dagger \bigr),
\label{eq:a_xi}\\
  a_\xi^\dagger &= \sum_\gamma 
    \bigl( \beta_{\xi\gamma}^* b_\gamma 
    + \alpha_{\xi\gamma}^* b_\gamma^\dagger 
    + \eta_{\xi\gamma}^* \tilde{b}_\gamma 
    + \zeta_{\xi\gamma}^* \tilde{b}_\gamma^\dagger \bigr),
\label{eq:a_xi-dag}
\end{align}
where $b_\gamma$ and $b_\gamma^\dagger$ are the operators in Eqs.~(\ref{eq:ann},~\ref{eq:cre}), $\xi$ is the label in which the frequency $\omega$ with respect to $t$ is traded with the frequency $\Omega$ associated with the infalling time, and $\alpha_{\xi\gamma}$, $\beta_{\xi\gamma}$, $\zeta_{\xi\gamma}$, and $\eta_{\xi\gamma}$ are the Bogoliubov coefficients calculable using the standard field theory method~\cite{Unruh:1976db,Israel:1976ur}.
The generator of the time evolution we are looking for would then be given by
\begin{equation}
  H = \sum_\xi \Omega a_\xi^\dagger a_\xi 
    + H_{\rm int}\bigl( a_\xi, a_\xi^\dagger \bigr).
\label{eq:H_eff}
\end{equation}
The question is if we can find microscopic operators that realize this and have appropriate properties to play the role of quantum operators in a semiclassical theory.

This was studied in Ref.~\cite{Nomura:2019dlz}, in which such operators were given.
The simplest possibility is to use Eq.~(\ref{eq:coarse}) in Eqs.~(\ref{eq:ann-m},~\ref{eq:cre-m}) to get
\begin{align}
  \tilde{b}_\gamma &= z\, \sum_n \sqrt{n_\gamma}\, e^{\frac{E_{n_-}+E_n}{2T_{\rm H}}}
    \sum_{i_{n_-} = 1}^{e^{S_{\rm bh}(M-E_{n_-})}} \sum_{j_n = 1}^{e^{S_{\rm bh}(M-E_n)}} \sum_{a = 1}^{e^{S_{\rm rad}}} \sum_{b = 1}^{e^{S_{\rm rad}}} 
    c_{n_- i_{n_-} a} c^*_{n j_n b}\, \ket{\psi^{(n_-)}_{i_{n_-}}} \ket{\phi_a} \bra{\psi^{(n)}_{j_n}} \bra{\phi_b},
\label{eq:ann-m-orig}\\
  \tilde{b}_\gamma^\dagger &= z\, \sum_n \sqrt{n_\gamma + 1}\, e^{\frac{E_{n_+}+E_n}{2T_{\rm H}}}
    \sum_{i_{n_+} = 1}^{e^{S_{\rm bh}(M-E_{n_+})}} \sum_{j_n = 1}^{e^{S_{\rm bh}(M-E_n)}} \sum_{a = 1}^{e^{S_{\rm rad}}} \sum_{b = 1}^{e^{S_{\rm rad}}} 
    c_{n_+ i_{n_+} a} c^*_{n j_n b}\, \ket{\psi^{(n_+)}_{i_{n_+}}} \ket{\phi_a} \bra{\psi^{(n)}_{j_n}} \bra{\phi_b},
\label{eq:cre-m-orig}
\end{align}
where $z = \sum_m e^{-E_m/T_{\rm H}}$, $n_\pm \equiv \{ n_\alpha \pm \delta_{\alpha\gamma} \}$, and $E_{n_\pm}$ are the energies of the hard mode states $\ket{\{ n_\alpha \pm \delta_{\alpha\gamma} \}}$ as measured in the asymptotic region.
Since $\tilde{b}_\gamma^\dagger$ decreases the energy, this ``intermediate'' operator can be defined only state dependently~\cite{Almheiri:2013hfa}.
However, because of the restriction to the hard modes, the final, physically relevant operators---appropriately smoothed $b_\gamma$, $b_\gamma^\dagger$, $a_\xi$, and $a_\xi^\dagger$ operators in the physical region---can be promoted to global, state-independent (linear) operators acting throughout the space of microstates~\cite{Nomura:2019dlz} as in Eq.~(\ref{eq:linear-op});%
\footnote{The non-injective nature of $\tilde{b}_\gamma^\dagger$ discussed in Ref.~\cite{Almheiri:2013hfa} leads to only exponentially suppressed effects in the description based on the global linear operators.}
for example, following Eq.~(\ref{eq:linear-op})
\begin{equation}
  a_\xi^\dagger = \sum_{A = 1}^{e^{S_{\rm bh}(M) + S_{\rm rad}}} a_\xi^{A\dagger},
\label{eq:cre-inf-global}
\end{equation}
where an appropriate smoothing is implied.
Here, $a_\xi^{A\dagger}$ is given by Eq.~(\ref{eq:a_xi-dag}) with $\tilde{b}_\gamma$ and $\tilde{b}_\gamma^\dagger$ replaced with $\tilde{b}^A_\gamma$ and $\tilde{b}_\gamma^{A\dagger}$, which are obtained by taking $c_{n i_n a}$'s in Eqs.~(\ref{eq:ann-m-orig},~\ref{eq:cre-m-orig}) to be those of a specific microstate, $c^A_{n i_n a}$, and $A$ runs over a set of orthogonal microstates.
Note that the operators in Eqs.~(\ref{eq:ann-m-orig},~\ref{eq:cre-m-orig}) involve both soft mode and radiation degrees of freedom.
This construction works regardless of the age of the black hole, i.e.\ both before and after the Page time.

If the black hole is younger, i.e.\ before the Page time, one can use the Petz map to obtain operators that act only on the soft modes
\begin{align}
  \tilde{b}_\gamma &= z\, e^{S_{\rm rad}} \sum_n \sqrt{n_\gamma}\, e^{\frac{E_{n_-} + E_n}{2 T_{\rm H}}} 
    \sum_{i_{n_-} = 1}^{e^{S_{\rm bh}(M-E_{n_-})}} \sum_{j_n = 1}^{e^{S_{\rm bh}(M-E_n)}} \sum_{a = 1}^{e^{S_{\rm rad}}} 
    c_{n_- i_{n_-} a} c_{n j_n a}^*\, \ket{\psi^{(n_-)}_{i_{n_-}}} \bra{\psi^{(n)}_{j_n}},
\label{eq:ann-av}\\
  \tilde{b}_\gamma^\dagger &= z\, e^{S_{\rm rad}} \sum_n \sqrt{n_\gamma + 1}\, e^{\frac{E_{n_+} + E_n}{2 T_{\rm H}}} 
    \sum_{i_{n_+} = 1}^{e^{S_{\rm bh}(M-E_{n_+})}} \sum_{j_n = 1}^{e^{S_{\rm bh}(M-E_n)}} \sum_{a = 1}^{e^{S_{\rm rad}}} 
    c_{n_+ i_{n_+} a} c_{n j_n a}^*\, \ket{\psi^{(n_+)}_{i_{n_+}}} \bra{\psi^{(n)}_{j_n}}.
\label{eq:cre-av}
\end{align}
We, however, cannot obtain analogous operators acting only on radiation~\cite{Nomura:2019dlz}.%
\footnote{The Petz map was used in Ref.~\cite{Penington:2019kki} to construct interior operators acting only on radiation.
This was possible because the analysis did not consider the energy constraint imposed on the black hole system, i.e.\ the hard and soft modes.
Indeed, if this constraint were not imposed, then the Petz map construction analogous to Eqs.~(\ref{eq:ann-av},~\ref{eq:cre-av}) would work to give operators acting only on radiation after the Page time~\cite{Nomura:2019dlz}.}

The existence of linear operators reproducing the correct semiclassical algebra throughout the space of microstates (up to the intrinsic ambiguity of order $e^{-E_{\rm exc}/T_{\rm H}}$) implies that there is a sector in the microscopic theory which encodes the experience of the object in near empty spacetime after it crosses the horizon.
This allows us to erect an effective theory of the interior at time $t$.
Since the effective theory is obtained by coarse-graining the region outside the zone (radiation), it describes only a limited spacetime region:\ the causal domain of the union of the zone and its mirror region on the spatial hypersurface at $t$ in the effective two-sided geometry.

\begin{figure}[t]
\begin{center}
  \includegraphics[height=8cm]{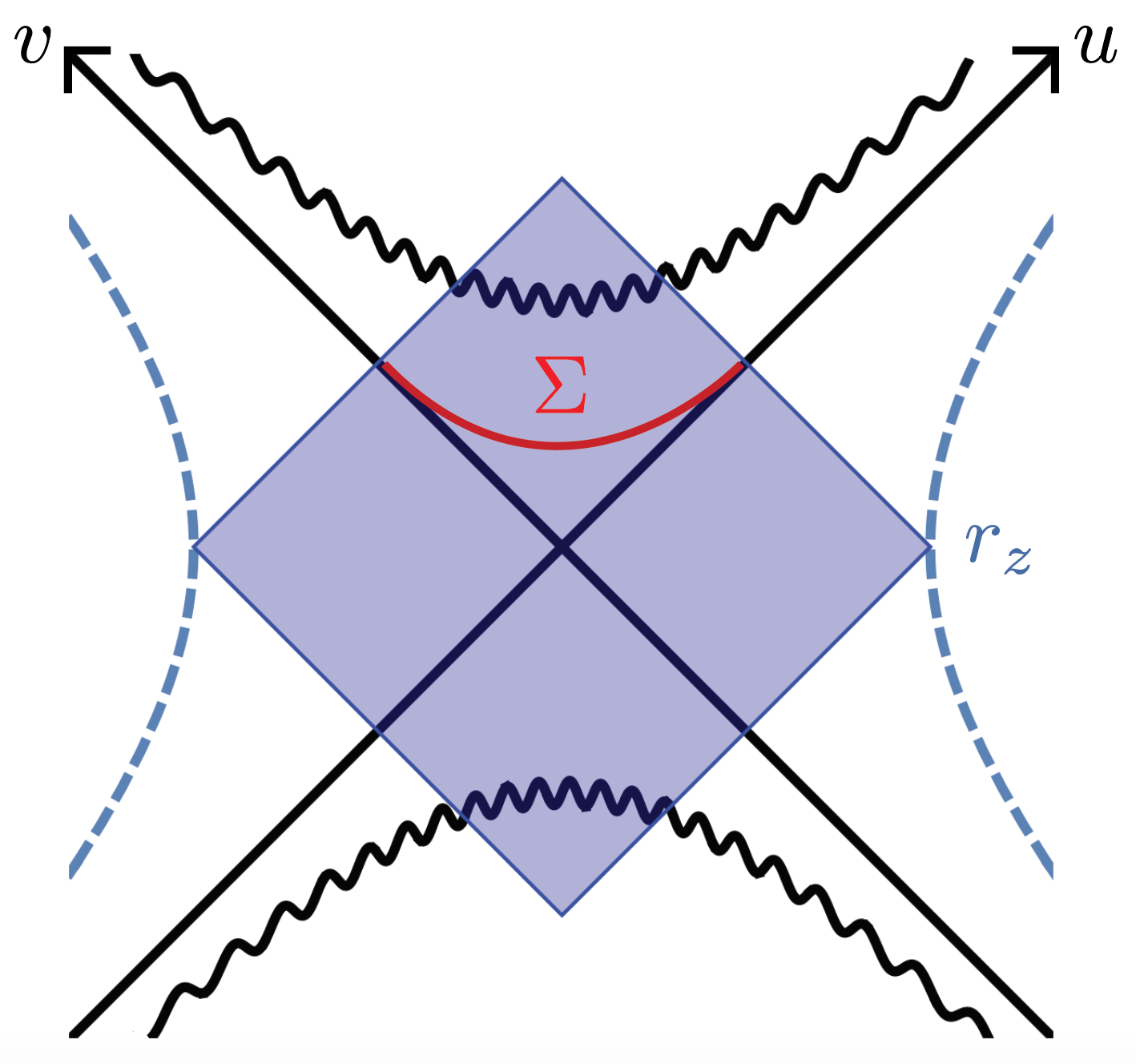}
\end{center}
\caption{Within the spacetime region described by the effect theory of the interior (diamond at the center), the interior hypersurface having the maximal volume ($\Sigma$ in red) is bounded by the codimension-2 surfaces given by the intersections of the horizon and future-directed light rays emitted from $r = r_{\rm z}$ and its mirror.
 The volume of this hypersurface is finite.}
\label{fig:volume}
\end{figure}
This limitation of the spacetime region solves the problem of an infinite volume.
The maximal interior volume one can consider is now that of hypersurfaces bounded by the codimension-2 surfaces given by the intersections of the horizon and future-directed light rays emitted from $r = r_{\rm z}$ and its mirror; see Fig.~\ref{fig:volume}.
Since the state of the black hole in the effective theory is given by the thermofield double form at the time when the effective theory is erected---regardless of the age of the original one-sided black hole---this volume is finite.
For a black hole in $(d+1)$-dimensional flat spacetime, for example, it is given by $V_{\rm max} \approx O(r_+^d)$, where $r_+$ is the horizon radius.
The amount of entropy of semiclassical matter one can place in this volume, without causing significant backreaction to the geometry, is indeed much smaller than the Bekenstein-Hawking entropy of the black hole.
A similar statement also applies to a large AdS black hole, where $V_{\rm max} \approx O(r_+^{d-1} l)$ with $l$ being the AdS radius.

The fact that an effective theory represents only a limited spacetime region implies that the picture of the whole interior, as described by general relativity, can be obtained only by using multiple effective theories erected at different times.
In the global gauge construction, this is manifested in the fact that seemingly independent interior states are not actually independent, as we have seen in Section~\ref{sec:ensemble}.

\subsection{Entanglement wedge reconstruction: reconstruction with time evolution}
\label{subsec:e-wedge}

There has been significant recent progress in understanding the interior of an evaporating black hole in the global gauge construction~\cite{Penington:2019npb,Almheiri:2019psf,Almheiri:2019hni,Penington:2019kki,Almheiri:2019qdq}.
The analyses employ either holographic entanglement wedge reconstruction~\cite{Czech:2012bh,Wall:2012uf,Headrick:2014cta,Jafferis:2015del,Dong:2016eik,Cotler:2017erl}, which builds on relations between bulk quantities and boundary quantum entanglement~\cite{Ryu:2006bv,Hubeny:2007xt,Faulkner:2013ana,Engelhardt:2014gca}, or Euclidean gravitational path integral including the effect of replica wormholes~\cite{Penington:2019kki,Almheiri:2019qdq}.
According to these analyses, operators acting on early radiation are sufficient to reconstruct a portion of the black hole interior after the Page time.
On the other hand, we have seen that our construction of the effective theory must involve soft mode degrees of freedom in addition to early radiation.
How can the relation between the two approaches be understood in the unitary gauge construction?

A key ingredient is the boundary time evolution.
In general, entanglement wedge reconstruction assumes that we know the time evolution operator of the boundary theory, in the models in AdS spacetime discussed in Refs.~\cite{Penington:2019npb,Almheiri:2019psf,Almheiri:2019hni,Penington:2019kki,Almheiri:2019qdq} the Hamiltonian of a system consisting of boundary conformal field theory and any auxiliary theory coupled to it.
In addition, in these models it is assumed that the information leaked from the boundary conformal field theory---representing the bulk with a black hole---to the auxiliary system---a system storing Hawking radiation---is effectively irreversible.
These conditions allow us to reconstruct a portion of interior spacetime given the state of radiation at some time after the Page time, using the boundary time evolution~\cite{Nomura:2019dlz}.

Let us discuss how this works in more detail.
It uses the fact that given the complete knowledge about radiation after the Page time $t_{\rm Page}$, the information that was fully scrambled into the black hole can be recovered from it once the appropriate amount of quanta emitted after the scrambling is added to it~\cite{Hayden:2007cs}.
Suppose we want to reconstruct the interior using the state of radiation at time $t$ ($> t_{\rm Page}$).
Suppose also that the state of the system at some time $t_{\rm w}$ ($t_{\rm Page} < t_{\rm w} < t$) takes the form of Eq.~(\ref{eq:sys-state}) with some hard modes corresponding to an infalling object being excited:
\begin{equation}
  f(\{ b_\gamma^\dagger \})\, \ket{\Psi(M)} = \sum_n \sum_{i_n = 1}^{e^{S_{\rm bh}(M-E_n)}} 
    \sum_{a = 1}^{e^{S_{\rm rad}}} d_{n i_n a} \ket{\{ n_\alpha \}} 
    \ket{\psi^{(n)}_{i_n}} \ket{\phi_a},
\label{eq:sys-tw}
\end{equation}
where the coefficients $d_{n i_n a}$ are different from those of the vacuum because of the excitations, $\{ d_{n i_n a} \} \neq \{ c_{n i_n a} \}$, and $f$ is a function specific to the object.
Our interest is to reproduce the fate of this object given the state of radiation at time $t$ ($> t_{\rm w}$).

We split the hard modes at time $t_{\rm w}$ into two classes:\ one that will (eventually) collide with the stretched horizon and be scrambled into the soft modes and the other that will leave the zone propagating into asymptotic space without being absorbed into the black hole.%
\footnote{The first class includes the modes which are initially outgoing but will be reflected back due to the potential barrier responsible for the graybody factor and absorbed into the stretched horizon.}
We label these two classes of modes by indices $\beta$ and $\gamma$, respectively
\begin{equation}
  \{ \alpha \} = \{ \beta \} + \{ \gamma \},
\end{equation}
and denote the occupation numbers of them by $\mu_\beta$ and $\nu_\gamma$:
\begin{equation}
  \{ n_\alpha \} = \left( \{\mu_\beta\}, \{\nu_\gamma\} \right).
\end{equation}
Note that the energies measured in the asymptotic region satisfy
\begin{equation}
  E_n = E_\mu + E_\nu,
\end{equation}
where $n = \{ n_\alpha \}$, $\mu = \{ \mu_\beta \}$, and $\nu = \{ \nu_\gamma \}$.

With this convention, the state in Eq.~(\ref{eq:sys-tw}) can be written as
\begin{equation}
  f(\{ b_\gamma^\dagger \})\, \ket{\Psi(M)} = \sum_\mu \sum_\nu \sum_{i_{(\mu,\nu)} = 1}^{e^{S_{\rm bh}(M-E_\mu-E_\nu)}} 
    \sum_{a = 1}^{e^{S_{\rm rad}}} d_{\mu\, \nu\, i_{(\mu,\nu)} a} \ket{\{ \mu_\beta \}} \ket{\{ \nu_\gamma \}} \bigl| \psi^{(\mu,\nu)}_{i_{(\mu,\nu)}} \bigr\rangle \ket{\phi_a}.
\label{eq:sys-tw-2}
\end{equation}
We now take $t - t_{\rm w}$ to be (sufficiently) larger than the scrambling time $t_{\rm scr}$~\cite{Hayden:2007cs,Sekino:2008he}:%
\footnote{More precisely, the explicit expression of the scrambling time in Eq.~(\ref{eq:scr}) is applicable to ingoing Eddington-Finkelstein time:\ $v_{\rm scr} \approx (1/2\pi T_{\rm H}) \ln S_{\rm bh}$~\cite{Penington:2019npb,Saraswat:2020zzf}.
The boundary scrambling time $t_{\rm scr}$ is in general smaller:\ $t_{\rm scr} \approx v_{\rm scr} - t_{\rm sig}$, where $t_{\rm sig}$ is the signal propagation time between the stretched horizon and the location where information is extracted, in this case the edge of the zone.
We ignore this subtlety here, since the difference between $v_{\rm scr}$ and $t_{\rm scr}$ is at most of $O(1)$:\ $1 < v_{\rm scr}/t_{\rm scr} \approx O(1)$.}
\begin{equation}
  t - t_{\rm w} > t_{\rm scr} \approx \frac{1}{2\pi T_{\rm H}} \ln S_{\rm bh}.
\label{eq:scr}
\end{equation}
The modes represented by $\beta$ are then scrambled into the soft modes by the time $t$, which we describe as follows.%
\footnote{As we see in Appendix~\ref{app:scr-prop}, the scrambling time is always of the order of or larger than the signal propagation time between the stretched horizon and the edge of the zone, so the modes reflected back by the potential barrier are also scrambled by $t > t_{\rm w} +O(t_{\rm scr})$.}
For each $\nu$
\begin{equation}
  \ket{\{ \mu_\beta \}} \bigl| \psi^{(\mu,\nu)}_{i_{(\mu,\nu)}} \bigr\rangle \quad\longrightarrow\quad U^{(\nu)} \ket{\{ \mu_\beta \}} \bigl| \psi^{(\mu,\nu)}_{i_{(\mu,\nu)}} \bigr\rangle \,\,\equiv\,\, \ket{\psi^{(\nu)}_{i_\nu}},
\label{eq:ev-scr}
\end{equation}
where $U^{(\nu)}$ is a unitary evolution operator, and the index $i_\nu$ depends on $\mu = \{ \mu_\beta \}$ and $i_{(\mu,\nu)}$
\begin{equation}
  i_\nu = i_\nu(\mu, i_{(\mu,\nu)}).
\label{eq:i_nu}
\end{equation}
This evolution can indeed be unitary because the change of the coarse-grained entropy is
\begin{equation}
  S_\mu + S_{\rm bh}(M-E_n) \quad\longrightarrow\quad S_{\rm bh}(M-E_n+E_\mu) \,\approx\, S_{\rm bh}(M-E_n) + \frac{E_\mu}{T_{\rm H}},
\end{equation}
so that the process increases the entropy as long as the Bekenstein bound, $S_\mu < E_\mu/T_{\rm H}$, is satisfied.
Some of the $\beta$ modes scrambled into the soft modes are further emitted to the asymptotic space as Hawking radiation, but we treat this separately later.

As we see in Appendix~\ref{app:scr-prop}, the scrambling time is always of the order of or larger than the signal propagation time between the stretched horizon and the edge of the zone
\begin{equation}
  t_{\rm scr} \gtrsim t_{\rm sig}.
\end{equation}
Thus, by the time $t$, the modes represented by $\nu$ are all emitted as radiation.
We describe this as
\begin{equation}
  \ket{\{ \nu_\gamma \}} \ket{\phi_a} \quad\longrightarrow\quad V^{(\nu,a)} \ket{\{ \nu_\gamma \}} \ket{\phi_a} \,\,\equiv\,\, \ket{\phi_{\nu,a}} 
\label{eq:ev-emit}
\end{equation}
for each $(\nu,a)$.

Substituting Eqs.~(\ref{eq:ev-scr}) and (\ref{eq:ev-emit}) into Eq.~(\ref{eq:sys-tw-2}), we find that the state (of the soft and far modes) at time $t$ becomes
\begin{equation}
  \ket{\Psi}_t \sim \sum_\nu \sideset{}{'}\sum_{i_\nu} \sum_{a = 1}^{e^{S_{\rm rad}}} d_{\nu i_\nu a} \ket{\psi^{(\nu)}_{i_\nu}}\ket{\phi_{\nu,a}},
\label{eq:sys-t}
\end{equation}
up to effects discussed below.
Note that $i_\nu$ is a function of $\mu$ and $i_{(\mu,\nu)}$, Eq.~(\ref{eq:i_nu}), so that the summation over $i_\nu$ is, in fact, the summations over $\mu$ and $i_{(\mu,\nu)}$
\begin{equation}
  \sideset{}{'}\sum_{i_\nu} = \sum_\mu \sum_{i_{(\mu,\nu)} = 1}^{e^{S_{\rm bh}(M-E_\mu-E_\nu)}}
\end{equation}
and we have defined $d_{\nu i_\nu a} \equiv d_{\mu\, \nu\, i_{(\mu,\nu)} a}$.
The actual state at $t$ contains hard modes at that time, which were a part of soft or far modes at time $t_{\rm w}$.
With the chaotic dynamics at the stretched horizon, the population of these modes is dictated by thermality, and the state takes the standard form of Eq.~(\ref{eq:sys-state}).
Since the number of hard modes is much smaller than that of soft modes, however, the effect of the appearance of these modes on the other sectors is minor, so we ignore it.

There is one important effect which we have not yet taken into account:\ conversion of soft modes into radiation outside the zone through Hawking emission.%
\footnote{This process occurs at the edge of the zone; see Refs.~\cite{Nomura:2018kia,Nomura:2014woa}.}
This is the effect that allows us to reconstruct the interior based only on radiation, through the Hayden-Preskill protocol~\cite{Hayden:2007cs}.
Notice that the soft mode states $\ket{\psi^{(\nu)}_{i_\nu}}$ and radiation states $\ket{\phi_{\nu,a}}$ in Eq.~(\ref{eq:sys-t}) contain the structure needed to reconstruct $\mu$ and $\nu$ and the degrees of freedom entangled with them.
In these states, let us separate the degrees of freedom, ${\cal I}_{\rm rec}$, needed for our reconstruction at $t$, and the rest, ${\cal I}_{\rm junk}$.
The fact that the Hayden-Preskill protocol works implies that the emission process can be written as
\begin{equation}
  \sum_\nu \sideset{}{'}\sum_{i_\nu} \sum_{a = 1}^{e^{S_{\rm rad}}} d_{\nu i_\nu a} \ket{\psi^{(\nu)}_{i_\nu}}\ket{\phi_{\nu,a}} \quad\longrightarrow\quad \sum_r \sum_j \sum_b c_{rb} c_{jb} \ket{\psi_j} \ket{\phi_{r,j,b}},
\label{eq:emission}
\end{equation}
where $r$ and $j$ are the indices for ${\cal I}_{\rm rec}$ and ${\cal I}_{\rm junk}$, respectively, and $b$ represents the radiation degrees of freedom that are not directly associated with $r$ or $j$.
The point is that the degrees of freedom represented by $r$ are now fully contained in radiation states.

This allows us to reconstruct the interior region that can be described by the effective theory erected at $t_{\rm w}$ on radiation at time $t$.
Specifically, denoting the unitary causing the evolution in Eq.~(\ref{eq:emission}) by $W$, we can reproduce the effect of acting $b_\gamma$ on the state at $t_{\rm w}$ by operators acting on the state at $t$
\begin{equation}
  {\cal B}_\gamma = W V U \,b_\gamma\, U^\dagger V^\dagger W^\dagger,
\end{equation}
and likewise for $b_\gamma^\dagger$:\ $({\cal B}_\gamma,b_\gamma) \rightarrow ({\cal B}_\gamma^\dagger,b_\gamma^\dagger)$.
Here, $U$ and $V$ are given by the unitaries in Eqs.~(\ref{eq:ev-scr},~\ref{eq:ev-emit}) as $U = \oplus_\nu U^{(\nu)}$ and $V = \oplus_{(\nu,a)} V^{(\nu,a)}$, where it is understood that $U^{(\nu)}$ and $V^{(\nu,a)}$ come with the projection operators onto the states with the corresponding values of $\nu$ and $(\nu,a)$.
Reflecting the fact that ${\cal I}_{\rm rec}$ can be fully accessed in radiation states, ${\cal B}_\gamma$ and ${\cal B}_\gamma^\dagger$ operate only on the radiation component when acted on a state at time $t$.

A similar construction also works for the infalling mode operators:
\begin{equation}
  {\cal A}_\xi = W V U \,a_\xi\, U^\dagger V^\dagger W^\dagger,
\end{equation}
and likewise for $a_\xi^\dagger$:\ $({\cal A}_\xi,a_\xi) \rightarrow ({\cal A}_\xi^\dagger,a_\xi^\dagger)$.
As discussed in Section~\ref{subsec:state-indep}, the Fock spaces obtained by acting ${\cal A}_\xi^\dagger$'s on different, orthogonal vacuum microstates are orthogonal up to the exponentially suppressed ambiguity.
We can therefore form global, approximately state-independent operators as in Eq.~(\ref{eq:linear-op}):
\begin{equation}
  \hat{A}_\xi = \sum_{A = 1}^{e^{S_{\rm bh} + S_{\rm rad}}} {\cal A}_\xi^{(A)} P_A,
\label{eq:global-rad}
\end{equation}
and similarly for $\hat{\cal A}_\xi^\dagger$.
Here, $A$ runs over a set of orthogonal states that were vacuum microstates at $t_{\rm w}$, and $P_A$ is the projection operator onto the Fock space built on $A$.
The fact that complete information about the black hole vacuum state is available in radiation after the Page time allows us to take $P_A$ to act only on the radiation component.
The global operators in Eq.~(\ref{eq:global-rad}) can thus be defined within the space of radiation states at time $t$.

\begin{figure}[t]
\begin{center}
  \includegraphics[height=8cm]{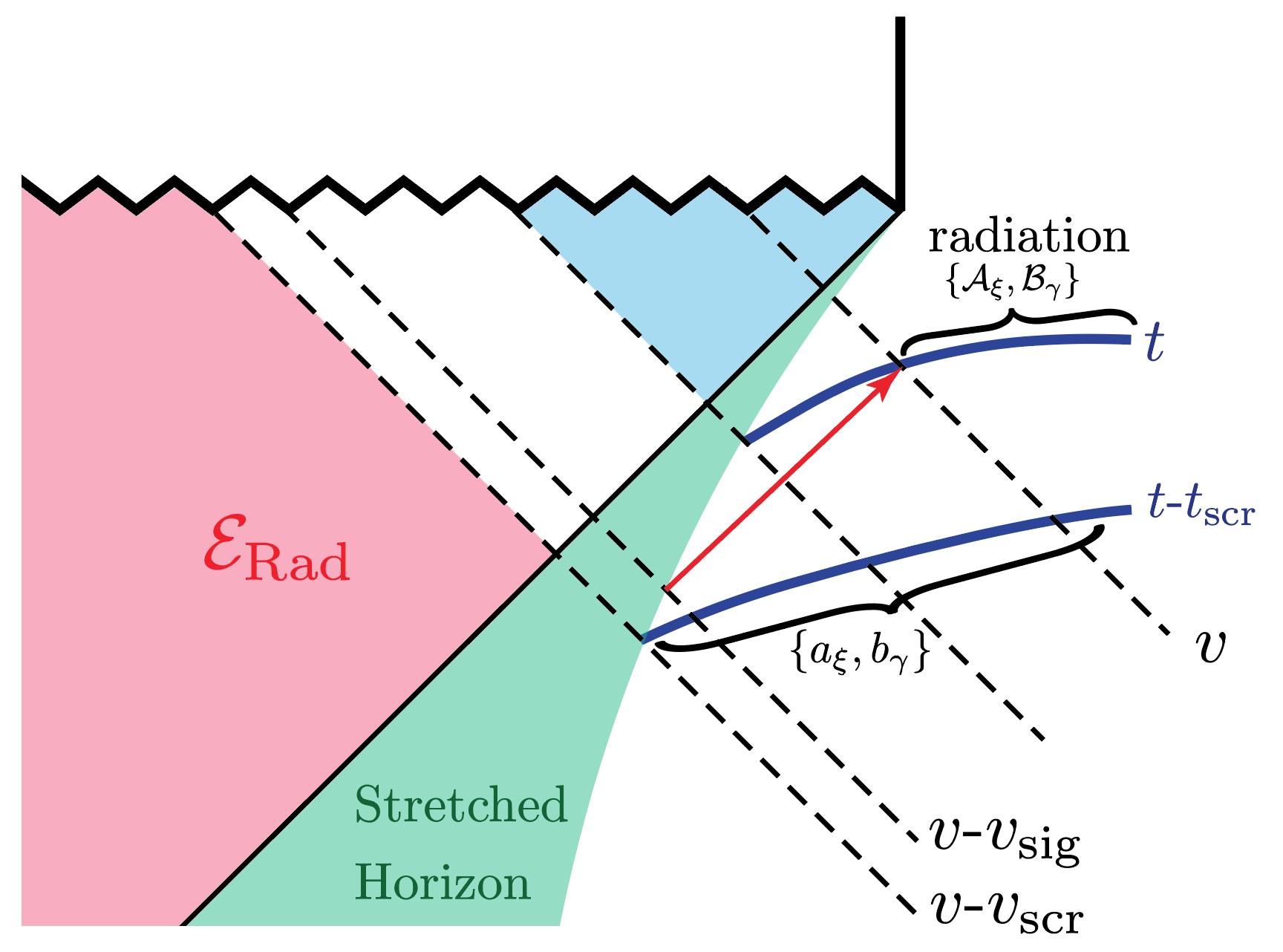}
\end{center}
\caption{Operators (${\cal A}_\xi$, ${\cal B}_\gamma$) acted on radiation states at time $t$ which have the same effect as operators of the effective interior theory ($a_\xi$, $b_\gamma$) erected at time $t - t_{\rm scr}$ (or earlier).
 This implies that one can reconstruct the entanglement wedge ${\cal E}_{\rm rad}$ from radiation states at time $t$.
 The blue shaded region is the interior region relevant to an object that is in the zone at time $t$ and falls into the black hole.
 The red arrow indicates a light signal propagating from the stretched horizon to the edge of the zone.}
\label{fig:wedge}
\end{figure}

The construction described above works as long as $t - t_{\rm w}$ is larger than a time of order the scrambling time $t_{\rm scr}$; more precisely,
\begin{equation}
  t - t_{\rm w} \,\gtrsim\, v_{\rm scr} - t_{\rm sig} \,\equiv\, t - t_{\rm w,max}.
\end{equation}
Here, $v_{\rm scr}$ is the ingoing Eddington-Finkelstein time needed to recover the information after an object hits the stretched horizon, while $t_{\rm sig}$ is the boundary time it takes for a signal to propagate from the stretched horizon to the location where the information is extracted (see Appendix~\ref{app:scr-prop}).
This implies that we can reconstruct the interior region that can be described by any effective theory erected before $t_{\rm w,max}$, using operators that act on radiation at time $t$.
This gives the entanglement wedge of the radiation obtained in Refs.~\cite{Penington:2019npb,Almheiri:2019psf,Almheiri:2019hni,Saraswat:2020zzf,Almheiri:2019psy,Rozali:2019day,Chen:2019uhq,Almheiri:2019yqk,Gautason:2020tmk,Anegawa:2020ezn,Hashimoto:2020cas,Hartman:2020swn,Hollowood:2020cou,Krishnan:2020oun}; see Fig.~\ref{fig:wedge}.%
\footnote{Here, we have focused on the interior portion of the entanglement wedge.
 (We have also ignored possible stretching inside the horizon.)
 Depending on the setup, e.g.\ how and where soft and hard modes are converted or extracted to radiation, the entanglement wedge may contain an exterior region as in Refs.~\cite{Almheiri:2019yqk,Almheiri:2019psy,Hollowood:2020cou,Chen:2020jvn}.
 This can occur because the relevant far mode degrees of freedom may be able to represent hard mode operators there, e.g.\ with the outgoing hard modes escaping the zone directly while ingoing hard modes first converted into soft modes (at the stretched horizon) and then escaping.}

While the entanglement wedge represents the spacetime region one can reconstruct from the radiation, the amount of information one can reconstruct, or the size of code subspace~\cite{Almheiri:2014lwa,Pastawski:2015qua,Harlow:2016vwg} one can erect, is not uniform throughout the entanglement wedge.%
\footnote{A similar issue was discussed in Refs.~\cite{Penington:2019npb,Hayden:2018khn} in a setup without the energy constraint.}
This feature inherits from the use of the Hayden-Preskill protocol---and more fundamentally the boundary time evolution---in the reconstruction.
Because of the scrambling and quantum error correcting nature of the black hole dynamics, one can arbitrarily choose {\it which} information ${\cal I}_{\rm rec}$ to reconstruct~\cite{Hayden:2007cs}.
However, there is an upper bound on the {\it amount} of information one can reconstruct, coming from the fact that in order to reconstruct an object carrying entropy $S_{\rm obj}$, one has to collect $O(S_{\rm obj})$ Hawking quanta, which takes a time of $O(S_{\rm obj}/T_{\rm H})$.
If $S_{\rm obj} \gtrsim \ln S_{\rm bh}$, this time is longer than the scrambling time $t_{\rm scr}$.

This implies that an object having entropy larger than $O(\ln S_{\rm bh})$ can be reconstructed only if it is located sufficiently in the past within the entanglement wedge.
Or equivalently, to reconstruct such an object which is absorbed by the stretched horizon at time $t_0$, one needs to use radiation at some time later than $t_0 + O(t_{\rm scr})$.
Specifically, to reconstruct an object carrying $S_{\rm obj}$ which enters the stretched horizon at time $t_0$, one needs to use the state of radiation at time $t$ with
\begin{equation}
  t - t_0 \gtrsim \begin{cases}
    \frac{1}{2\pi T_{\rm H}} \ln S_{\rm bh} & \mbox{for } S_{\rm obj} \lesssim \ln S_{\rm bh} \\
    \frac{1}{2\pi T_{\rm H}} S_{\rm obj} & \mbox{for } S_{\rm obj} \gtrsim \ln S_{\rm bh}.
  \end{cases}
\end{equation}
This structure is not visible if we only compute the location of the entanglement wedge by the quantum extremization procedure.

As can be seen in the discussion above, there are certain drawbacks in the entanglement wedge reconstruction using only radiation degrees of freedom.
First, since the reconstruction involves time evolution backward in time, the expressions for the bulk operators in terms of boundary operators acting on radiation are highly complicated, and the reconstructed operators are extremely fragile; i.e., a small deformation of the boundary  operators destroys the success of the reconstruction.
Furthermore, since the reconstructed spacetime region is that described by effective theories erected more than the scrambling time in the past, the reconstructed operators do not describe the interior region that is relevant to the fate of an object located in the zone region at the time when the state is given.
In order to erect an effective theory that is capable of describing future evolution of such an object, we need to use operators that act both on the soft modes and radiation as discussed in Section~\ref{subsec:without-ev}.

\section{Conclusions}
\label{sec:concl}

The information problem of black holes boils down to the tension between the unitarity of the evolution as viewed from a distance (e.g.\ the S-matrix) and the existence of near empty spacetime inside the horizon.
In this paper, we have promoted the idea that there are two complementary descriptions of an evaporating black hole at the quantum level---the global gauge and unitary gauge constructions---and that we can understand aspects that are mysterious in one description better by using the other description.

In the global gauge construction, one begins with the global spacetime of general relativity, so that the existence of interior spacetime is given.
The challenge then is to understand the unitarity of the evolution, as illustrated by the original, ``naive'' calculation by Hawking~\cite{Hawking:1976ra}.
This issue has recently been addressed successfully, at least in simple models of gravity in low dimensions, by the discovery of new saddles in the gravitational path integral~\cite{Penington:2019kki,Almheiri:2019qdq}.
Entanglement wedge reconstruction was employed to understand which degrees of freedom carry information about the interior~\cite{Penington:2019npb,Almheiri:2019psf,Almheiri:2019hni}.
We note, however, that entanglement wedge reconstruction does not explain how the interior emerges; the existence of the interior is assumed.
It simply says that if the interior exists (and it does by construction), then the degrees of freedom whose entanglement wedge contains a portion of the interior can be used to reconstruct it by the protocol of Ref.~\cite{Hayden:2007cs}.

In the unitary gauge construction, one instead begins with a manifestly unitary description.
This description corresponds to the picture of a black hole as viewed from a distance.
Since the (stretched) horizon in this picture behaves as a physical membrane~\cite{Susskind:1993if}, there is no problem in understanding unitarity; the only difference from a regular material surface in this respect is that we do not (yet) know the explicit microscopic dynamics of the surface.
The fact that the stretched horizon behaves as a material surface, however, brings the question of the interior~\cite{Almheiri:2012rt}; what is special about the stretched horizon, allowing an object to fall through it without even noticing it?
This issue was addressed in Refs.~\cite{Nomura:2018kia,Nomura:2019qps,Nomura:2019dlz} by providing explicit constructions of operators that make the fate of the fallen object manifest.
Special dynamical properties of the horizon---maximally chaotic, fast scrambling, and universal---are crucial for the success of the constructions, singling out the stretched horizon.
The constructed operators, which include the generator of infalling time evolution, are linear throughout the space of microstates and satisfy the algebra of standard quantum field theory in near empty spacetime, up to exponentially small ambiguities.
It is therefore plausible that the infalling object indeed experiences the smooth horizon.

The fact that the two, seemingly very different constructions lead to the same physical conclusions implies that they are different descriptions of the same system.
It is reasonable to expect that this is a manifestation of enormous nonperturbative gauge redundancies of a gravitational theory discussed in Refs.~\cite{Marolf:2020xie,McNamara:2020uza,Jafferis:2017tiu}.
In this context, the appearance of a horizon seems to be indicative of the situation where the nonperturbative redundancies play an important, leading role in understanding the correct physics.
And as such, similar issues are also expected to arise in inflationary cosmology~\cite{Nomura:2019qps,Nomura:2011dt,Bousso:2011up}.
It is interesting to understand the full implications of these nonperturbative redundancies, which would indeed have a fundamental importance in quantum cosmology.

\section*{Acknowledgments}

We are grateful to Adam Bouland, Chitraang Murdia, Pratik Rath, and Douglas Stanford for discussions on this and related topics.
This work was supported in part by the Department of Energy, Office of Science, Office of High Energy Physics under contract DE-AC02-05CH11231 and award DE-SC0019380.

\appendix

\section{Stretched Horizon and Soft Modes}
\label{app:sh-sm}

In this appendix, we show that
\begin{itemize}
\item
The stretched horizon---the location at which the classical description of spacetime breaks down---can be defined either as a surface on which the local Hawking temperature becomes the inverse string length or a surface whose proper distance from the mathematical horizon is the string length; and these two definitions agree.
\item
The mass and entropy of a black hole can be viewed as being carried by the soft modes---modes whose frequencies are of order the local Hawking temperature or smaller.
\end{itemize}
We focus on a spherically symmetric black hole in $d+1$ dimensions ($d \geq 3$) which is not near extremal.

The metric is given by
\begin{equation}
  ds^2 = -f(r) dt^2 + \frac{1}{f(r)} dr^2 + r^2 d\Omega_{d-1}^2.
\end{equation}
The mathematical (outer) horizon, $r = r_+$, is then given by the largest real positive root of $f(r)$:
\begin{equation}
  f(r_+) = 0.
\end{equation}
The temperature and entropy of the black hole are given by
\begin{equation}
  T_{\rm H} = \frac{f'(r_+)}{4\pi}
\label{eq:T_H}
\end{equation}
and
\begin{equation}
  S_{\rm bh} = \frac{r_+^{d-1}}{4 G_{\rm N}}\, {\rm vol}(\Omega_{d-1}),
\label{eq:S_BH}
\end{equation}
respectively, where $G_{\rm N}$ is the Newton constant, and ${\rm vol}(\Omega_{d-1}) = 2\pi^{d/2}/\Gamma(d/2)$ is the volume of the $(d-1)$-dimensional unit sphere.
The local temperature is defined as
\begin{equation}
  T_{\rm loc}(r) \,=\, \frac{T_{\rm H}}{\sqrt{f(r)}} \,=\, \frac{f'(r_+)}{4\pi\sqrt{f(r)}}.
\end{equation}

\subsection{Stretched horizon}
\label{subsec:strethced}

We can define the stretched horizon, $r = r_{\rm s}$, using the condition on the local temperature:
\begin{equation}
  T_{\rm loc}(r_{\rm s}) \approx \frac{1}{2\pi l_{\rm s}},
\label{eq:str-def-1}
\end{equation}
where $l_{\rm s}$ is the string length. By writing $r_{\rm s} = r_+ + \delta r$, we obtain
\begin{equation}
  T_{\rm loc}(r_+ + \delta r) \,=\, \frac{1}{4\pi} \frac{f'(r_+)}{\sqrt{f(r_+ + \delta r)}} \,=\, \frac{1}{4\pi} \sqrt{\frac{f'(r_+)}{\delta r}}
\end{equation}
at the leading order in $\delta r$, so that
\begin{equation}
  \delta r \,\equiv\, r_{\rm s} - r_+ \,\approx\, \frac{f'(r_+)\, l_{\rm s}^2}{4}.
\label{eq:delta-r}
\end{equation}

Alternatively, we may define the stretched horizon as a surface which is a string length away from the mathematical horizon:
\begin{equation}
  \int_{r_+}^{r_{\rm s}} \frac{dr}{\sqrt{f(r)}} \approx l_{\rm s}.
\end{equation}
In this case, the leading order expansion gives
\begin{equation}
  \int_{r_+}^{r_{\rm s}} \frac{dr}{\sqrt{f(r)}} 
  \,=\, \int_{r_+}^{r_+ + \delta r}\!\! \frac{dr}{\sqrt{f'(r)\, (r - r_+)}} 
  \,=\, \frac{2 \sqrt{\delta r}}{\sqrt{f'(r_+)}}.
\end{equation}
This again leads to Eq.~(\ref{eq:delta-r}).

\subsection{Soft modes as black hole microstates}
\label{subsec:soft}

Let us integrate the entropy and local energy densities of the soft modes from the stretched horizon, $r = r_{\rm s}$, to the edge of the black hole region, $r = r_{\rm z}$. This gives
\begin{alignat}{5}
  S &\sim N \int_{r_{\rm s}}^{r_{\rm z}}\! T_{\rm loc}(r)^d 
    \frac{r^{d-1} dr}{\sqrt{f(r)}} 
  &&\sim \frac{N\, T_{\rm H}^d\, r_+^{d-1}}
    {f'(r_+)^{\frac{d+1}{2}} \delta r^{\frac{d-1}{2}}} 
  &&\sim \frac{N r_+^{d-1}}{l_{\rm s}^{d-1}} 
  &&\sim \frac{r_+^{d-1}}{G_{\rm N}},
\label{eq:BH-S}\\
  E &\sim N \int_{r_{\rm s}}^{r_{\rm z}}\! T_{\rm loc}(r)^{d+1} 
    \frac{r^{d-1} dr}{\sqrt{f(r)}} 
  &&\sim \frac{N\, T_{\rm H}^{d+1}\, r_+^{d-1}}
    {f'(r_+)^{\frac{d+2}{2}} \delta r^{\frac{d}{2}}} 
  &&\sim \frac{T_{\rm H}\, r_+^{d-1}}{G_{\rm N} \sqrt{f'(r_+) \delta r}} 
  &&\sim \frac{M}{\sqrt{-g_{tt}(r_{\rm s})}},
\label{eq:BH-E}
\end{alignat}
where we have assumed that the integrals are dominated at $r = r_{\rm s}$ (which is justified for a realistic black hole) and used Eqs.~(\ref{eq:T_H}) and (\ref{eq:delta-r}); $N$ is the number of low energy species below the string scale, which satisfies the relation
\begin{equation}
  \frac{l_{\rm s}^{d-1}}{N} \sim G_{\rm N},
\label{eq:ls-GN}
\end{equation}
and $M$ in the last expression in Eq.~(\ref{eq:BH-E}) is the mass of the black hole.

The location of the edge of the zone, $r = r_{\rm z}$, is determined by analyzing the effective potential $V_\ell(r)$ appearing in the scalar equation of motion
\begin{equation}
  \left[ -\frac{d^2}{(dr^*)^2} + V_\ell(r) - \omega^2 \right] \chi_{\omega L_d} = 0,
\end{equation}
where $r^*$ is the tortoise coordinate, $dr^* = dr/f(r)$, and $\chi_{\omega L_d}$ are the modes of a scalar field $\varphi$ defined by
\begin{equation}
  \varphi(t,r,\Omega) = \frac{1}{r^{\frac{d-1}{2}}} \sum_{\omega, L_d} 
    \chi_{\omega L_d}(r)\, Y_{L_d}(\Omega)\, e^{-i \omega t}.
\end{equation}
The explicit form of the potential is given by
\begin{equation}
  V_\ell(r) = \frac{d-1}{2}\frac{f(r)^2}{r^2} \left( r\frac{f'(r)}{f(r)} + \frac{d-3}{2} \right) 
    + \ell(\ell+d-2) \frac{f(r)}{r^2}.
\label{eq:V_ell}
\end{equation}
This will be analyzed later to determine $r_{\rm z}$.

The results in Eqs.~(\ref{eq:BH-S},~\ref{eq:BH-E}) show that the entropy and energy carried by the soft modes correctly reproduce, parametrically, the entropy and energy of the black hole, with the latter being measured at the stretched horizon, where most of the soft modes reside. This allows us to view that the entropy and energy of the black hole are indeed carried by the soft modes.

\subsection{Examples}
\label{subsec:ex}

We now discuss explicit examples, applying the results so far. This elucidates some issues that were not discussed explicitly, e.g.\ how we should choose the edge of the black hole region $r_{\rm z}$.

The metric of $(d+1)$-dimensional AdS Schwarzschild spacetime ($d \geq 3$) is given by
\begin{equation}
  f(r) = 1 - \frac{r_+^{d-2}}{r^{d-2}} + \frac{r^2}{l^2} - \frac{r_+^d}{l^2 r^{d-2}},
\end{equation}
where $l$ is the AdS radius. The black hole mass is given by
\begin{equation}
  M = (d-1) \left( 1 + \frac{r_+^2}{l^2} \right) 
    \frac{r_+^{d-2}}{16 \pi G_{\rm N}}\, {\rm vol}(\Omega_{d-1}).
\end{equation}
The temperature and entropy of the black hole are given by
\begin{equation}
  T_{\rm H} = \frac{d\, r_+^2 + (d-2) l^2}{4\pi r_+ l^2},
\qquad
  S_{\rm bh} = \frac{r_+^{d-1}}{4 G_{\rm N}}\, {\rm vol}(\Omega_{d-1}).
\end{equation}

\subsubsection{Flat space (or small AdS) black hole}

This case is obtained by taking the $l/r \rightarrow \infty$ limit, leading to
\begin{equation}
  f(r) = 1 - \frac{r_+^{d-2}}{r^{d-2}},
\qquad
  M = (d-1) \frac{r_+^{d-2}}{16 \pi G_{\rm N}}\, {\rm vol}(\Omega_{d-1}),
\label{eq:generic-flat-1}
\end{equation}
\begin{equation}
  T_{\rm H} = \frac{d-2}{4\pi r_+},
\qquad
  S_{\rm bh} = \frac{r_+^{d-1}}{4 G_{\rm N}}\, {\rm vol}(\Omega_{d-1}).
\label{eq:generic-flat-2}
\end{equation}
The stretched horizon is located at
\begin{equation}
  r_{\rm s} - r_+ \approx \frac{d-2}{4} \frac{l_{\rm s}^2}{r_+}.
\label{eq:generic-flat-3}
\end{equation}

The effective potential in Eq.~(\ref{eq:V_ell}) has a peak around $r - r_+ \approx O(r_+)$, suggesting that we should take $r_{\rm z}$ near this peak:%
\footnote{We can define the zone radius $r_{\rm z}$ precisely as the radius satisfying $\lim_{\ell \rightarrow \infty} V'_\ell(r_{\rm z}) = 0$ and $\lim_{\ell \rightarrow \infty} V''_\ell(r_{\rm z}) < 0$, although the precise value of $r_{\rm z}$ is not important anyway.}
\begin{equation}
  r_{\rm z} \approx \left(\frac{d}{2}\right)^{\frac{1}{d-2}} r_+.
\label{eq:generic-flat-4}
\end{equation}
We can indeed view the surface $r = r_{\rm z}$ as the boundary between the near black hole and asymptotically flat regions.

\subsubsection{Large AdS black hole}

For a large AdS black hole $r_+ \gg l$,
\begin{equation}
  f(r) = \frac{r^2}{l^2} - \frac{r_+^d}{l^2 r^{d-2}},
\qquad
  M = (d-1) \frac{r_+^d}{16 \pi G_{\rm N} l^2}\, {\rm vol}(\Omega_{d-1}),
\label{eq:generic-large-1}
\end{equation}
\begin{equation}
  T_{\rm H} = \frac{d\, r_+}{4\pi l^2},
\qquad
  S_{\rm bh} = \frac{r_+^{d-1}}{4 G_{\rm N}}\, {\rm vol}(\Omega_{d-1}).
\label{eq:generic-large-2}
\end{equation}
The stretched horizon is at
\begin{equation}
  r_{\rm s} - r_+ \approx \frac{d}{4} \frac{r_+ l_{\rm s}^2}{l^2}.
\label{eq:generic-large-3}
\end{equation}

In this case, the effective potential in Eq.~(\ref{eq:V_ell}) monotonically increases with $r$, so we can take
\begin{equation}
  r_{\rm z} \approx \infty.
\label{eq:generic-large-4}
\end{equation}
In fact, the integrals in Eqs.~(\ref{eq:BH-S},~\ref{eq:BH-E}) converge with $r_{\rm z} \rightarrow \infty$ because of the AdS nature. We can, therefore, view the entire AdS system as the near black hole region. This is consistent with the fact that in order to make a large AdS black hole evaporate, we need to couple the AdS spacetime with an auxiliary ``bath'' system, as in Refs.~\cite{Penington:2019npb,Almheiri:2019psf,Rocha:2008fe}.

\section{Scrambling Time vs Signal Propagation Time}
\label{app:scr-prop}

In this appendix, we show that in all cases under consideration the scrambling time is larger, or of the same order, compared with the signal propagation time.

\subsection{General consideration}

As in Appendix~\ref{app:sh-sm}, we consider a spherically symmetric black hole in $d+1$ dimensions ($d \geq 3$):
\begin{equation}
  ds^2 = -f(r) dt^2 + \frac{1}{f(r)} dr^2 + r^2 d\Omega_{d-1}^2.
\end{equation}
We also use the ingoing Eddington-Finkelstein coordinates, whose metric is
\begin{align}
  ds^2 = -f(r)dv^2 +2dv dr + r^2d\Omega^2_{d-1},
\label{eq:ingoing-EF}
\end{align}
where
\begin{equation}
  v = t + r^*,
\end{equation}
and $r^*$ is the tortoise coordinate defined by
\begin{equation}
  dr^* = \frac{dr}{f(r)}.
\end{equation}

We define the signal propagation time as the time it takes for a radially outgoing light ray to propagate from the stretched horizon $r_{\rm s}$ to the radius $r_{\rm e}$ at which information is extracted (in the case of spontaneous Hawking emission, $r_{\rm e}$ is the edge of the zone $r_{\rm z}$).
Here, $r_{\rm s}$ and $r_{\rm z}$ are defined in Appendix~\ref{app:sh-sm}.
We denote the signal propagation time in boundary time $t$ by
\begin{equation}
  t_{\rm sig} = \varDelta r^*,
\end{equation}
where $\varDelta r^*$ is the distance between $r_{\rm s}$ and $r_{\rm e}$ in the tortoise coordinate.
In ingoing Eddington-Finkelstein time, this quantity is given by
\begin{equation}
  v_{\rm sig} \,=\, t_{\rm sig} + \varDelta r^* \,=\, 2 t_{\rm sig}.
\label{eq:v_sig}
\end{equation}
On the other hand, the calculation of the entanglement wedge of radiation gives us the scrambling time $v_{\rm scr}$~\cite{Penington:2019npb,Saraswat:2020zzf}.
This is the length of time in the ingoing Eddington-Finkelstein coordinates between the time at which an object hits the stretched horizon and the earliest time at which its information can be extracted at radius $r_{\rm e}$.
The corresponding quantity in boundary time is thus
\begin{equation}
  t_{\rm scr} = v_{\rm scr} - \varDelta r^*.
\label{eq:t_scr}
\end{equation}

From Eqs.~(\ref{eq:v_sig},~\ref{eq:t_scr}), we obtain the expression
\begin{equation}
  t_{\rm scr} = v_{\rm scr} - v_{\rm sig} + t_{\rm sig}.
\end{equation}
Since $t_{\rm scr} > 0$, this leads to
\begin{equation}
  v_{\rm scr} - v_{\rm sig} > -t_{\rm sig}.
\end{equation}
In this appendix, we demonstrate that the scrambling and signal propagation times satisfy stronger inequality
\begin{equation}
  v_{\rm scr} \geq v_{\rm sig},
\end{equation}
as expected from causality.
We explicitly show this up to fractional corrections of order $1/(\ln S_{\rm bh})$.
Here, $S_{\rm bh}$ is the entropy of the black hole, and we assume $\ln S_{\rm bh} \gg 1$.
Using Eqs.~(\ref{eq:v_sig},~\ref{eq:t_scr}), the inequality can be translated into
\begin{equation}
  t_{\rm scr} \geq t_{\rm sig}.
\end{equation}

\subsection{General analysis}

Here we try to make as much progress as possible in a model independent manner.

\subsubsection{Signal propagation time}

The signal propagation time $v_{\rm sig}$ in the Eddington-Finkelstein coordinates is given from Eq.~(\ref{eq:ingoing-EF}) as
\begin{equation}
  v_{\rm sig} = 2\int_{r_{\rm s}}^{r_z}\frac{dr}{f(r)}.
\label{eq:sig-integ}
\end{equation}
For a flat space or small AdS black hole, the integral in Eq.~(\ref{eq:sig-integ}) is dominated by the near horizon region and so we can approximate it as
\begin{equation}
  v_{\rm sig} \,\approx 2\, \int_{\delta r}^{r_{\rm e}-r_+}\! \frac{dx}{f'(r_+)\, x} \,=\, \frac{1}{2\pi T_{\rm H}} \ln\frac{r_{\rm e} - r_+}{\delta r} \,\approx\, \frac{1}{2\pi T_{\rm H}} \left[ \ln\frac{r_{\rm e} - r_+}{l_{\rm s}^2 T_{\rm H}} + O(1) \right].
\label{eq:v_sig-small}
\end{equation}
For a large AdS black hole, where $r_{\rm z} = \infty$, the integral in Eq.~(\ref{eq:sig-integ}) saturates above $x \sim r_+$ and hence
\begin{equation}
  v_{\rm sig} \approx \begin{cases}
    \frac{1}{2\pi T_{\rm H}} \ln\frac{r_{\rm e} - r_+}{\delta r} 
    \,\approx\, \frac{1}{2\pi T_{\rm H}} \ln\frac{r_{\rm e} - r_+}{l_{\rm s}^2 T_{\rm H}} &
    \mbox{for } r_{\rm e} - r_+ \lesssim r_+
  \\
    \frac{1}{2\pi T_{\rm H}} \ln\frac{r_+}{\delta r} 
    \,\approx\, \frac{1}{2\pi T_{\rm H}} \ln\frac{r_+}{l_{\rm s}^2 T_{\rm H}} &
    \mbox{for } r_{\rm e} - r_+ \gtrsim r_+
  \end{cases}
\label{eq:v_sig-large}
\end{equation}
at the leading logarithmic level.
Here, we have assumed $l \gg l_{\rm s}$.

\subsubsection{Scrambling time}

The scrambling time for a flat space (or small AdS) black hole was calculated in Ref.~\cite{Penington:2019npb} under the setup that Hawking radiation is extracted at some radius $r_{\rm e}$ in the zone using all the modes available there.
The resulting expression for the scrambling time is%
\footnote{Here and below, we assume that the spatial dimension $d$ is not extremely large.}
\begin{equation}
  v_{\rm scr} = \frac{1}{2\pi T_{\rm H}} \left[ \ln\frac{S_{\rm bh}}{c_{\rm evap}} + O(1) \right].
\label{eq:v_scr-Pen}
\end{equation}
The coefficient $c_{\rm evap}$ in Eq.~(\ref{eq:v_scr-Pen}) represents the number of available species when the system is viewed in 2~dimensions through the Kaluza-Klein decomposition, which can be evaluated as
\begin{equation}
  c_{\rm evap} \sim N_{\rm e} \sum_{\ell = 0}^{\ell_{\rm max}} \ell^{d-2} \sim N_{\rm e} \ell_{\rm max}^{d-1},
\label{eq:c_evap}
\end{equation}
where we have used the fact that the number of independent states with a fixed orbital quantum number $\ell$ goes as $\ell^{d-2}$ in $d+1$ dimensions, and $N_{\rm e}$ is the number of species available at $r = r_{\rm e}$ in the original ($d+1$)-dimensional theory.
The maximal value of $\ell$ in Eq.~(\ref{eq:c_evap}) is determined by the condition that the energy cost of angular momentum is not larger than the Hawking temperature in the original theory as
\begin{equation}
  \ell_{\rm max} \,\sim\, \frac{r_+ T_{\rm H}}{\sqrt{f(r_{\rm e})}} \,\sim\, \sqrt{\frac{r_+^2 T_{\rm H}}{r_{\rm e} - r_+}}.
\label{eq:l_max}
\end{equation}

The scrambling time for a large AdS black hole was computed in Ref.~\cite{Saraswat:2020zzf}.
The expression found there can be written in the form%
\footnote{We disagree with the statement in Ref.~\cite{Saraswat:2020zzf} that the leading term of $v_{\rm scr}$ is not given by the logarithm of the entropy of the horizon of the black hole.
 We find that it can still be written in the standard form in terms of the temperature and the entropy of the entire horizon.}
\begin{equation}
  v_{\rm scr} = \frac{1}{2\pi T_{\rm H}} \left[ \ln\frac{S_{\rm bh}}{c_{\rm evap}} + O(1) \right]
\label{eq:v_scr-large-1}
\end{equation}
for $r_{\rm e} - r_+ \lesssim r_+$.
The scrambling time for $r_{\rm e} - r_+ \gg r_+$ was not calculated in Ref.~\cite{Saraswat:2020zzf}, but we expect that it is still given by the fundamental expression in Eq.~(\ref{eq:v_scr-large-1}) as long as there is an available angular momentum mode, i.e.\ $c_{\rm evap}/N_{\rm e} \gtrsim O(1)$.
(If this is not the case, our result below would apply only to $r_{\rm e} - r_+ \lesssim r_+$.)
For large values of $r_{\rm e}$ for which $c_{\rm evap}/N_{\rm e} \lesssim O(1)$, the information extraction occurs through freely propagating radiation, so we expect that $v_{\rm scr}$ is constant there (at the leading logarithmic level) because of the AdS nature of spacetime.

\subsection{Comparison}

We now compare the scrambling time to the signal propagation time.
All the expressions below are given at the leading logarithmic level.

\subsubsection{Flat space (small AdS) black hole}

Consider a flat space (or small AdS) black hole.
Properties of this black hole are given in Eqs.~(\ref{eq:generic-flat-1}~--~\ref{eq:generic-flat-4}).

The signal propagation time and the scrambling time are given by Eqs.~(\ref{eq:v_sig-small}) and (\ref{eq:v_scr-Pen}) as
\begin{equation}
  v_{\rm sig} \,\approx\, \frac{1}{2\pi T_{\rm H}} \ln\frac{r_{\rm e} - r_+}{f'(r_+) l_{\rm s}^2} \,\approx\, \frac{2r_+}{d-2} \ln\frac{r_+(r_{\rm e}-r_+)}{l_{\rm s}^2}
\end{equation}
and
\begin{equation}
  v_{\rm scr} \,\approx\, \frac{1}{2\pi T_{\rm H}} \ln\frac{S_{\rm bh}}{c_{\rm evap}} \,\approx\, \frac{(d-1)r_+}{d-2} \ln\frac{r_+(r_{\rm e}-r_+)}{l_{\rm s}^2},
\end{equation}
respectively.
Taking the ratio, we obtain
\begin{equation}
  \frac{v_{\rm scr}}{v_{\rm sig}} \approx \frac{d-1}{2}.
\label{eq:ratio}
\end{equation}
We indeed find $v_{\rm scr}/v_{\rm sig} \geq 1$ for $d \geq 3$.

\subsubsection{Large AdS black hole}

We now discuss a large AdS black hole.
Properties of this black hole are given in Eqs.~(\ref{eq:generic-large-1}~--~\ref{eq:generic-large-4}).

The signal propagation time is given by Eq.~(\ref{eq:v_sig-large}) as
\begin{equation}
  v_{\rm sig} \approx \begin{cases}
    \frac{1}{2\pi T_{\rm H}} \ln\frac{r_{\rm e} - r_+}{f'(r_+) l_{\rm s}^2} 
    \,\approx\, \frac{2 l^2}{d\, r_+} \ln\frac{(r_{\rm e}-r_+) l^2}{r_+ l_{\rm s}^2} &
    \mbox{for } r_{\rm e} - r_+ \lesssim r_+
  \\
    \frac{1}{2\pi T_{\rm H}} \ln\frac{r_+}{f'(r_+) l_{\rm s}^2} 
    \,\approx\, \frac{2 l^2}{d\, r_+} \ln\frac{l^2}{l_{\rm s}^2} &
    \mbox{for } r_{\rm e} \gtrsim r_+.
  \end{cases}
\end{equation}
The scrambling time is given by Eq.~(\ref{eq:v_scr-large-1}) with the consideration about the value of $c_{\rm evap}/N_{\rm e}$:
\begin{equation}
  v_{\rm scr} \approx \begin{cases}
    \frac{(d-1) l^2}{d\, r_+} \ln\frac{(r_{\rm e}-r_+) l^2}{r_+ l_{\rm s}^2} &
    \mbox{for } r_{\rm e} \lesssim \frac{r_+^3}{l^2}
  \\
    \frac{(d-1) l^2}{d\, r_+} \ln\frac{r_+^2}{l_{\rm s}^2} &
    \mbox{for } r_{\rm e} \gtrsim \frac{r_+^3}{l^2}.
  \end{cases}
\end{equation}
By taking the ratio, we obtain
\begin{equation}
  \frac{v_{\rm scr}}{v_{\rm sig}} \approx \begin{cases}
    \frac{d-1}{2} & \mbox{for } r_{\rm e} - r_+ \lesssim r_+
  \\
    \frac{d-1}{2} \frac{\ln[(r_{\rm e}-r_+) l^2/r_+ l_{\rm s}^2]}
      {\ln(l^2/l_{\rm s}^2)} 
    \,\gtrsim\, \frac{d-1}{2} 
    & \mbox{for } r_+ \lesssim r_{\rm e} \lesssim \frac{r_+^3}{l^2}
  \\
    \frac{d-1}{2} \frac{\ln(r_+^2/l_{\rm s}^2)}{\ln(l^2/l_{\rm s}^2)} 
    \,\gtrsim\, \frac{d-1}{2} 
    & \mbox{for } r_{\rm e} \gtrsim \frac{r_+^3}{l^2},
  \end{cases}
\end{equation}
and hence $v_{\rm scr}/v_{\rm sig} \geq 1$ for $d \geq 3$.

\end{document}